\documentclass[twoside,journal]{IEEEtran}
\usepackage[T1]{fontenc}
\usepackage[utf8]{inputenc}
\usepackage{xcolor}
\usepackage{verbatim}
\usepackage{float}
\usepackage{multirow}
\usepackage{amsmath}
\usepackage{amssymb}
\usepackage{graphicx}
\usepackage{adjustbox}
\usepackage[unicode=true,
 bookmarks=false,
 breaklinks=false,pdfborder={0 0 1},backref=false,colorlinks=false]
 {hyperref}
\hypersetup{
 colorlinks,citecolor=black,urlcolor=black,linkcolor=black}

\makeatletter

\providecommand{\tabularnewline}{\\}
\floatstyle{ruled}
\newfloat{algorithm}{tbp}{loa}
\providecommand{\algorithmname}{Algorithm}
\floatname{algorithm}{\protect\algorithmname}

\let\oldforeign@language\foreign@language
\DeclareRobustCommand{\foreign@language}[1]{%
  \lowercase{\oldforeign@language{#1}}}

\floatstyle{ruled}
\newfloat{algorithm}{tbp}{loa}
\providecommand{\algorithmname}{Algorithm}
\floatname{algorithm}{\protect\algorithmname}

\usepackage{balance}
\usepackage{url}
\usepackage{cite}

\usepackage{fancyhdr}
\usepackage{amsfonts}
\usepackage{fixmath}
\usepackage{nicefrac}
\usepackage{resizegather}
\resizegathersetup{warningthreshold=1} 

\DeclareMathOperator*{\argmax}{arg\,max}

\usepackage{mdframed}
\newmdenv[ 
topline=false,
leftline=false,
rightline=false,
bottomline=false,
]{txtbox}

\usepackage{ragged2e}
\usepackage{booktabs}
\usepackage{array}
\usepackage{url}

\newcommand{\ie}{\textit{i.e.}}
\newcommand{\eg}{\textit{e.g.}}
\newcommand{\Tau}{\mathcal{T}}

\newcolumntype{P}[1]{>{\centering\arraybackslash}p{#1}}
\newcolumntype{M}[1]{>{\centering\arraybackslash}m{#1}}
\newcolumntype{L}[1]{>{\raggedright\let\newline\\\arraybackslash\hspace{0pt}}m{#1}}
\newcolumntype{C}[1]{>{\centering\let\newline\\\arraybackslash\hspace{0pt}}m{#1}}
\newcolumntype{R}[1]{>{\raggedleft\let\newline\\\arraybackslash\hspace{0pt}}m{#1}}
\usepackage{multirow}

\usepackage{algorithm}
\usepackage{algpseudocode}
\algnewcommand\algorithmicforeach{\textbf{for each}}
\algdef{S}[FOR]{ForEach}[1]{\algorithmicforeach\ #1\ \algorithmicdo}
\usepackage{etoolbox}
\usepackage{tikz}
\errorcontextlines\maxdimen
\usepackage{resizegather}

\usepackage{xcolor}
\newtheorem{theorem}{Theorem}\newtheorem{lemma}{Lemma}\newtheorem{proposition}[theorem]{Proposition}
\newtheorem{remark}{Remark}\newtheorem{definition}{Definition}

\newtheorem{example}{Example}

\makeatother

\begin{document}
\vspace*{\fill}

\begin{minipage}{1.0\textwidth}
	$\copyright$ \the\year{} IEEE.  Personal use of this material is permitted.  Permission from IEEE must be obtained for all other uses, in any current or future media, including reprinting/republishing this material for advertising or promotional purposes, creating new collective works, for resale or redistribution to servers or lists, or reuse of any copyrighted component of this work in other works.
\end{minipage}

\vspace*{\fill}
\thispagestyle{empty}
\newpage
\title{Multi-Objective Multi-Agent Planning for Discovering and Tracking
Multiple Mobile Objects}
\author{Hoa Van Nguyen, Ba-Ngu Vo, Ba-Tuong Vo, Hamid Rezatofighi and Damith
C. Ranasinghe \thanks{Acknowledgement: This work is supported by the Australian Research
Council under Projects LP160101177, LP200301507, and FT210100506.} \thanks{Hoa Van Nguyen, Ba-Tuong Vo and Ba-Ngu Vo are with the Department
of Electrical and Computer Engineering, Curtin University, Bentley,
WA 6102, Australia (e-mail: {hoa.v.nguyen,ba-tuong.vo,ba-ngu.vo}@curtin.edu.au).}\thanks{Hamid Rezatofighi is with the Department of Data Science \& AI, Monash
University, Clayton VIC 3800, Australia (e-mail: hamid.rezatofighi@monash.edu).}\thanks{Damith C. Ranasinghe is with the School of Computer Science, The University
of Adelaide, SA 5005, Australia (e-mail: damith.ranasinghe@adelaide.edu.au).} }
\markboth{PREPRINT: IEEE Trans on Signal Processing - DOI:10.1109/TSP.2024.3423755}{Nguyen \MakeLowercase{\textit{et. al.}}: Multi-Objective Multi-Agent
Planning for Discovering and Tracking Multiple Mobile Objects}
\maketitle
\setcounter{page}{1}
\begin{abstract}
We consider the online planning problem for a team of agents to discover
and track an unknown and time-varying number of moving objects from
onboard sensor measurements with uncertain measurement-object origins.
Since the onboard sensors have limited field-of-views, the usual
planning strategy based solely on either tracking detected objects
or discovering unseen objects is inadequate. To address this, we formulate
a new information-based multi-objective multi-agent control problem, cast
as a partially observable Markov decision process (POMDP). The resulting
multi-agent planning problem is exponentially complex due to the unknown
data association between objects and multi-sensor measurements; hence,
computing an optimal control action is intractable. We prove that
the proposed multi-objective value function is a monotone submodular
set function, which admits low-cost suboptimal solutions via greedy
search with a tight optimality bound. The resulting planning algorithm
has a \textit{linear} complexity in the number of objects and measurements
across the sensors, and \textit{quadratic} in the number of agents.
We demonstrate the proposed solution via a series of numerical experiments
with a real-world dataset. 
\end{abstract}


\IEEEpeerreviewmaketitle{}

\section{Introduction}

Recent advancements in robotics have inspired applications that use
low-cost mobile sensors (\eg, drones), ranging from vision-based
surveillance, threat detection via source localisation, search and
rescue, to wildlife monitoring~\cite{hoa2019jofr}. At the heart
of these applications is Multi-Object Tracking (MOT)---the task of
estimating an unknown and time-varying number of objects and their
trajectories from sensor measurements with unknown\textit{ data association}
(\ie, unknown measurement-to-object origin). Further, MOT is fundamental
for autonomous operation as it provides awareness of the dynamic environment
in which the agents operate. Although a single agent can be tasked
with MOT, such a system is limited by observability, computing resources,
and energy. Using multi-agents alleviates these problems, improves
synergy, and affords robustness to failures. Realising this potential
requires the multi-agents to collaborate and operate autonomously.

\begin{figure}[!tb]
\centering \includegraphics[width=0.3\textwidth]{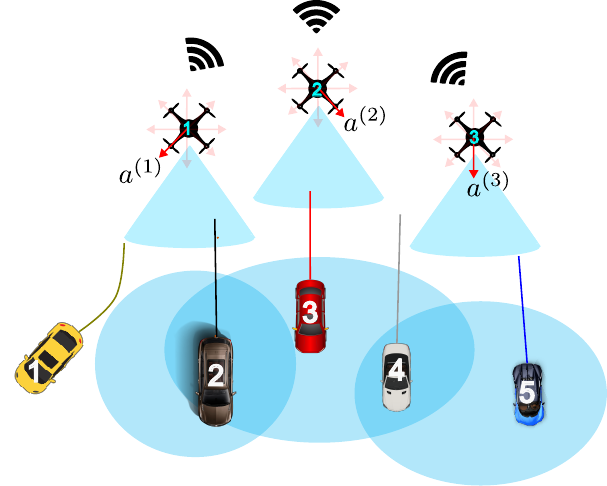}
\vspace{-0.3cm}
 \caption{An unmanned aerial vehicles (UAV) team tracking multiple vehicles
with limited FoV sensors and unknown measurements-to-objects associations.}
 \vspace{-0.5cm}
 \label{fig:target_applications} 
\end{figure}

In this work, we consider the challenging problem of coordinating
multiple limited field-of-view (FoV) agents to \textit{simultaneously}
seek \textit{undetected objects} and track \textit{detected objects}
(see Fig.~\ref{fig:target_applications}). Tracking involves estimating
the trajectories of the objects and maintaining their provisional
identities or labels~\cite{blackman1999design}. Trajectories are
important for capturing the behaviour of the objects while their labels
provide the means for distinguishing individual trajectories and for
human/machine users to communicate information on the relevant trajectories.
A single agent with an on-board sensor (\eg, a camera, a radar) invariably
has a limited FoV~\cite{farmani2014optimal}, and hence, only observes
part of the scene at any given time. As a result, a team of multiple
agents is often implemented to improve coverage of the surveillance
area~\cite{li2018robust,li2019computationally,gostar2020centralized,ong2020bayesian,wang2022multi,panicker2020tracking}.
However, MOT with multiple limited-FoV sensors still encountered several
challenges, such as occlusions, missed detections, false alarms, identity
switches and track fragmentation~\cite{hoa2021distributed,wang2023centralized}.

Discovering undetected objects and tracking detected objects are two
competing objectives due to the limited FoVs of the sensors 
and the random appearance/disappearance of objects.
On one hand, following only detected objects to track them accurately
means that many undetected objects could be missed. On the other hand,
leaving detected objects to explore unseen regions for undetected
objects will lead to track loss. Thus, the problem of seeking \textit{undetected
objects} and tracking \textit{detected objects} is a multi-objective
optimisation problem.

Even for standard state-space models, where the system state is a
finite-dimensional vector, multi-agent planning with multiple competing objectives is challenging due to complex interactions between agents
resulting in combinatorial optimisation problems \cite{welikala2022new}.
In MOT, where the system state (and measurement) is a set of vectors,
the problem is further complicated due to: \textit{i)} the unknown
and time-varying number of trajectories; \textit{ii)} missing and
false detections; and \textit{iii)} unknown data association (measurement-to-object
origins) \cite{mahler2014advances}. Most critically for real-world
applications, multi-agent control actions must be computed online
and in a timely manner.

Model predictive control (MPC) is an effective approach to stochastic
control and is widely used \cite{garcia1989model},
compared to meta-heuristics (or bio-inspired) techniques such as genetic
algorithms and particle swarm optimisation \cite{roberge2012comparison},
which are expensive for real-time applications in dynamic environments
such as MOT. The MPC problem can be cast as a partially observed Markov
decision process (POMDP), which has been gaining significant interest
as a real-time planning approach~\cite{clempner2019observer}. The
cooperation problem amongst agents can be formulated as a decentralised
POMDP (Dec-POMDP), whose exact solutions are NEXP-hard \cite{bernstein2002complexity}.
Moreover, for multi-agent Dec-POMDP, the action and observation space
grows exponentially with the number of agents \cite{bernstein2002complexity},
and hence unsuitable for real-time applications. This
intractability arises because the distributed agents do not necessarily
have the same formation states, and computing the optimal action requires
accounting for all possible observation histories and action sequences of the agents. In contrast,
all agents in a centralised POMDP have the \textit{same} information state (from the central node), which drastically reduces the complexity of action selection. Therefore, centralised Multi-agent POMDP~\cite{Messias2011} offers a more tractable alternative for coordinating multiple agents~\cite{dames2017detecting,Reza2}, and is adopted in our work. 

Computing optimal actions for MOT in a POMDP requires a suitable framework
that provides a multi-object density for the information state. Amongst
various MOT algorithms, we adopt the labelled random finite set (RFS)
framework because it provides a multi-object filtering density that
contains all information about the current set of trajectories and
accommodates its time-varying nature. Single-agent planning with RFS filters has been studied with an 
unlimited FoV~\cite{ristic2010sensor,hoang2014sensor} and extended
to multi-agent planning using distributed fusion \cite{Reza2}.
However, these multi-agent planning methods are only suitable when
the agents have an unlimited detection range~\cite{hoa2021distributed}.
A multi-agent POMDP with an RFS filter was proposed in \cite{dames2017detecting}
for searching and localisation, but not tracking.

\begin{figure}[!tb]
\centering \includegraphics[width=0.4\textwidth]{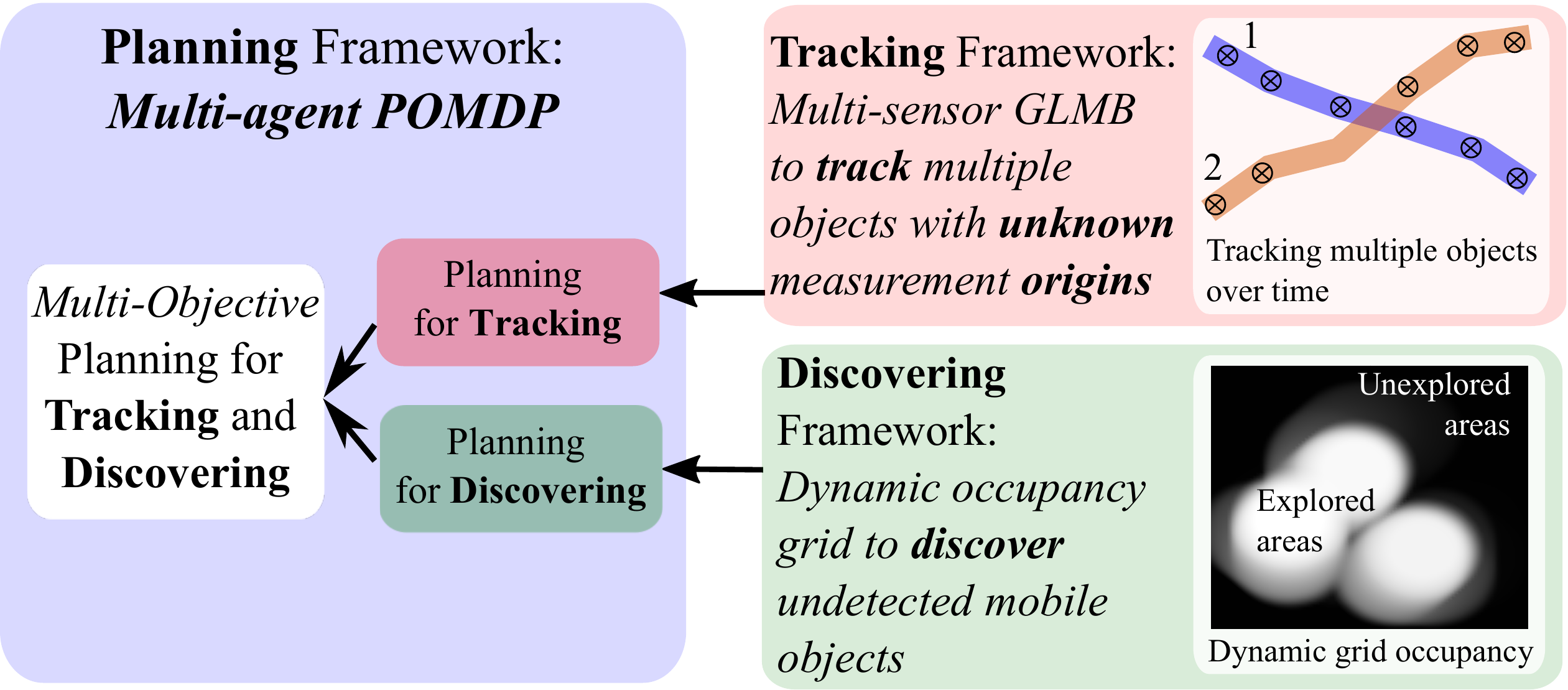}
\vspace{-0.3cm}
 \caption{A schematic of the proposed 
 planning approach. }
\vspace{-0.3cm}
 \label{fig:overall_planning_framework} 
\end{figure}

To achieve competing discovery and tracking objectives, we propose
a POMDP with a multi-objective value function consisting of information
gains for both detected and undetected objects, as illustrated in
Fig.~\ref{fig:overall_planning_framework}. A simple solution for
competing objective functions is to weigh and add them together. However,
meaningful weighting parameters are difficult to determine. A multi-objective
optimisation approach naturally provides a meaningful trade-off between
competing objectives. To the best of our knowledge, multi-agent planning
for MOT with a multi-objective value function has not been investigated.
The first multi-objective POMDP with a localisation RFS filter was
proposed in \cite{zhu2019multi} for a sensor selection problem.

For MOT and the information state, we use the multi-sensor Generalised
Labelled Multi-Bernoulli (MS-GLMB) filter~\cite{vo2019multisensor},
which exhibits \textit{linear} computational complexity in the total
number of measurements across the sensors. For path planning, we use
the centralised \textit{Multi-Agent POMDP} (MPOMDP) framework~\cite{Messias2011}
with the GLMB filter. Further, for the value function, we use the
differential entropy~\cite[pp.243]{cover2012elements} of the Labelled
Multi-Bernoulli (LMB) density~\cite{reuter2014lmb} that matches
the first moment of the information state. In particular, we derive
an analytic expression for this differential entropy, which can be
computed with \textit{linear} complexity in the number of hypothesised
objects, and show that the resulting multi-objective value function
is monotone submodular. Consequently, this enables us to exploit low-cost
greedy search to solve the optimal control problem with tight optimality
bound. In summary, the main contributions are: 
\begin{itemize}
\item[\textbf{\textit{i)}}] A multi-objective POMDP formulation for multi-agent planning to search
and track multiple objects with unknown object-to-measurement origins. 
\item[\textbf{\textit{ii)}}] The concept of differential entropy and mutual information for RFS, and an analytic expression for the differential entropy of LMBs with \textit{linear} complexity in the number of objects.
\item[\textbf{\textit{iii)}}] An efficient multi-agent path planning algorithm with \textit{linear} complexity in the number of objects and quadratic in the number of agents, using differential-entropy-based and occupancy-grid-based value functions. 
\end{itemize}
The proposed method is evaluated using the \textbf{CRAWDAD taxi dataset}~\cite{piorkowski2009crawdad}
on multi-UAVs control for searching and tracking an unknown and time-varying
number of taxis in downtown San Francisco. For performance benchmarking,
our preliminary work in ~\cite{hoa2020aaai} is used as the ideal
or best-case scenario where data association is known.

The remainder of this paper is organised as follows. Section~\ref{sec:background}
provides relevant background for our MOT and planning framework. Section~\ref{sec:planning}
presents our proposed multi-objective value function for the control
of the multi-agent to simultaneously track and discover objects. Section~\ref{sec:real_dataset}
evaluates our proposed method on a real-world dataset. Section~\ref{sec:conclusion}
summarises our contributions and discusses future directions.

\section{Background\label{sec:background}}

This section presents the problem statement and necessary
background on MPOMDP and MS-GLMB filtering. We use notation in accordance
with \cite{vo2013glmb}. Lowercase letters (\eg, $x,\boldsymbol{x}$)
denote vectors, uppercase letters (\eg, $X,\boldsymbol{X}$) denote
finite sets, while spaces are denoted by blackboard uppercase letters
(\eg, $\mathbb{X},\mathbb{L}$). Vectors augmented with labels, finite
sets of labelled vectors, and the corresponding spaces are bolded
(\eg, $\boldsymbol{x,X}$, $\pmb{\mathbb{X}},\pmb{\mathbb{U}}$)
to distinguish them from unlabelled ones. For a given set $S$, $|S|$
$,1_{S}(\cdot)$, $\delta_{S}[\cdot]$ denote, respectively, the cardinality,
indicator function, and (generalised) Kronecker delta function, of
$S$ ($\delta_{S}[X]=1$, if $X=S$, and zero otherwise). For a function
$f$, its multi-object exponential $f^{X}$ is defined as $\prod_{x\in X}f(x)$,
with $f^{\emptyset}=0$. The inner product $\int f(x)g(x)dx$ is written
as $\langle f,g\rangle$, while the superscript $^{\dagger}$ denotes
the transpose of a vector/matrix. For compactness, the subscript for
time $k$ is omitted, and subscripts for times $k-1$ and $k+1$ are
abbreviated `$-$' and `$+$'. A list of symbols
is provided in Table~\ref{tab:parameters}.

\begin{table}[!tb]
\centering  \caption{List of Symbols}
\vspace{-0.2cm}
 \label{tab:parameters} %
\begin{tabular}{|l|l|}
\hline 
\textbf{Symbols} & \textbf{Description}\tabularnewline
\hline 
$\mathcal{N}$  & finite set of agent labels \tabularnewline
$\mathcal{G}$  & Gaussian distribution\tabularnewline
$H$ & look-ahead horizon \tabularnewline
\hline 
$\mathbb{X}$  & unlabelled single-object state space \tabularnewline
$\mathbb{L}$  & label space \tabularnewline
$\pmb{\mathbb{X}}\triangleq\mathbb{X}\times\mathbb{L}$  & labelled single-object state space \tabularnewline
$\pmb{\mathcal{X}}$  & partition matroid of $\pmb{\mathbb{X}}$ \tabularnewline
$\boldsymbol{X} \in \pmb{\mathcal{X}} $  & labelled multi-object state \tabularnewline
$\mathbb{U}^{(n)}$  & state space of agent $n$ \tabularnewline
$u^{(n)} \in \mathbb{U}^{(n)} $  & state of agent $n$ \tabularnewline
\hline 
$\mathbb{Z}^{(n)}$  & observation space of agent $n$ \tabularnewline
$\pmb{\mathbb{Z}}\triangleq\uplus_{n\in\mathcal{N}}(\mathbb{Z}^{(n)}\times\{n\})$  & common observation space \tabularnewline
$\pmb{\mathcal{Z}}$  & class of finite subsets of $\pmb{\mathbb{Z}}$ 
\tabularnewline
$\boldsymbol{Z} \in \pmb{\mathcal{Z}} $  & multi-agent multi-object observation \tabularnewline
\hline 
$\mathbb{A}^{(n)}$  & control action space of agent $n$ \tabularnewline
$\mathbb{A}=\mathbb{A}^{(1)}\times...\times\mathbb{A}^{(|\mathcal{N}|)}$  & multi-agent action space \tabularnewline
$\pmb{\mathbb{A}}\triangleq\uplus_{n\in\mathcal{N}}(\mathbb{A}^{(n)}\times\{n\})$  & common action space \tabularnewline
$\pmb{\mathcal{A}}$  & partition matroid of $\pmb{\mathbb{A}}$ \tabularnewline
$a=(a^{(1)},...,a^{(|\mathcal{N}|)})$  & multi-agent action \tabularnewline
\hline 
$\boldsymbol{\varrho}$  & immediate reward function \tabularnewline
$h$  & differential entropy function\tabularnewline
$V_{1}$  & tracking value function \tabularnewline
$V_{2}$  & discovery value function \tabularnewline
$V_{mo}$  & multi-objective value function \tabularnewline
\hline 
$\kappa^{(i)}$  & occupancy grid cell $i$ \tabularnewline
$O^{(i)}$  & Bernoulli random variable of cell $i$ \tabularnewline
$Y^{(i)}$  & binary observation of cell $i$\tabularnewline
\hline 
\end{tabular} 
\end{table}

\subsection{Problem statement}

Consider a team of agents, equipped with limited-FoV sensors that
record detections with unknown\textit{ data association} (measurement-to-object
origins), monitoring an unknown and time-varying number of objects.
The agents can self-localise and communicate (\eg, sending observations)
to a central node that determines/issues control actions~\cite{dames2017detecting}.
For each agent $n\mathcal{\in N}=\{1,\dots,|\mathcal{N}|\}$, its
state space\footnote{The measurement errors of agents' internal actuator-sensor are assumed
negligible, thus the agent state is known.}, (discrete) control action space, and observation space, are denoted
respectively by $\mathbb{U}^{(n)}$, $\mathbb{A}^{(n)}$, and $\mathbb{Z}^{(n)}$.
We define the \textit{common observation space} for all agents as
$\pmb{\mathbb{Z}}\triangleq\uplus_{n\in\mathcal{N}}\big(\mathbb{Z}^{(n)}\times\{n\}\big)$.

Each object is described by a labelled state $\boldsymbol{x}=(x,\ell)\in\pmb{\mathbb{X}}\triangleq\mathbb{X}\times\mathbb{L}$,
consisting of a time-varying \textit{attribute} $x\in\mathbb{X}$,
and a time-invariant discrete \textit{label} $\ell\in\mathbb{L}$.
To capture the time-varying number of trajectories, the \textit{multi-object
state} at any given time is represented by the set $\boldsymbol{X}$
of distinctly labelled states of the objects \cite{vo2013glmb}, \cite{vo2019multiscan}.
Thus, the \textit{multi-object state space} $\pmb{\mathcal{X}}$ is
the \textit{partition matroid} of $\pmb{\mathbb{X}}$. A partition
matroid of a space $\pmb{\mathbb{S}}=\mathbb{S}\times\mathbb{I}$,
with $\mathbb{I}$ a discrete set, is the class of subsets of $\pmb{\mathbb{S}}$,
defined by~\cite{corah2018distributed}: 
\begin{align*}
\pmb{\mathcal{S}}\triangleq\{\boldsymbol{S}\subseteq\pmb{\mathbb{S}}:|\boldsymbol{S}\cap(\mathbb{S}\times\{i\})|\leq1,~\forall i\in\mathbb{I}\}.
\end{align*}
Each member of the partition matroid $\pmb{\mathcal{S}}$ has distinct
indices.

Our goal is to coordinate the team of agents to search and track a
time-varying and unknown number of moving objects. The main challenges
are: \textit{{i)}} misdetection of objects due to the limited FoVs;
\textit{ii)} lack of prior information about the changing number of
objects and their locations; \textit{iii)} unknown data association;
and \textit{iv)} the computational resources required to determine
multi-agent control actions in a timely manner. \vspace{-0.4cm}

\subsection{MPOMDP for Multi-Object Tracking}

A random finite set (RFS) model~\cite{mahler2007statistical,liu2018structure}
succinctly captures the random nature of the multi-object states and
offers a suitable concept of multi-object density for the POMDP's
\textit{information state}, encapsulating all current
knowledge about the system based on the observation history. Specifically,
\textit{an RFS of a space $\mathbb{X}$ is a random variable taking
values in the class 
of finite subsets of} $\mathbb{X}$,
and is commonly characterised by Mahler's \textit{multi-object density}
\cite{mahler2007statistical} (see also subsection~\ref{subsec:Object-DiffEnt}).
Since the elements of a multi-object state are distinctly labelled,
we model the multi-object state as a \textit{labelled RFS}--an RFS
of $\pmb{\mathbb{X}}$ such that instantiations have distinct labels
\cite{vo2013glmb}, i.e., a random variable taking values in the partition
matroid $\pmb{\mathcal{X}}$.

Conceptually, an MPOMDP is similar to a single-agent
POMDP, but incorporates joint models of all agents and communications
for sharing observations and/or information~\cite{Messias2011}.
This work explores a centralised approach\footnote{The communication traffic between agents and a central node is minimal if employing detection-based observations.} where agents send observations
to a central node responsible for coordinating the team via the control
actions. The multi-object MPOMDP considered in this work can be characterised
by the tuple $(\mathcal{N},H,\pmb{\mathcal{X}},\pmb{\mathcal{Z}},\mathcal{\mathbb{A}},\boldsymbol{f},\boldsymbol{g},\boldsymbol{\varrho})$,
where 
\begin{itemize}
\item $\mathcal{N}$ is the finite set of agent labels; 
\item $H$ is the look-ahead time horizon; 
\item $\pmb{\mathcal{X}}$ is the multi-object state space (partition matroid
of $\pmb{\mathbb{X}}$); 
\item $\pmb{\mathcal{Z}}$ is the multi-object observation space (class
of finite subsets of $\pmb{\mathbb{Z}}$); 
\item $\mathbb{A}=\mathbb{A}^{(1)}\times...\times\mathbb{A}^{(|\mathcal{N}|)}$
is multi-agent action space; 
\item $\boldsymbol{f}(\boldsymbol{X}|\boldsymbol{X}_{-},a_{-})$ is the
transition density for $\boldsymbol{X},\boldsymbol{X}_{-}\in\pmb{\mathcal{X}}$,
and action $a_{-}\in\mathbb{A}$ has been taken; 
\item $\boldsymbol{g}(\boldsymbol{Z}|\boldsymbol{X},a_{-})$ is the likelihood
of observation $\boldsymbol{Z}\in\pmb{\mathcal{Z}}$ given $\boldsymbol{X}\in\pmb{\mathcal{X}}$,
and action $a_{-}\in\mathbb{A}$ has been taken; 
\item $\boldsymbol{\varrho}(\boldsymbol{Z},\boldsymbol{X},a_{-})$ is the
immediate reward for observing $\boldsymbol{Z}\in\pmb{\mathcal{Z}}$
from $\boldsymbol{X}\in\pmb{\mathcal{X}}$, after action $a_{-}\in\mathbb{A}$
has been taken. 
\end{itemize}
The agents are coordinated at each time $k$ via the multi-agent action
$a=(a^{(1)},...,a^{(|\mathcal{N}|)})\in\mathbb{A}$, which prescribes
a state for each agent $n$ at time $k+1$, where it collects the
multi-object observation ${Z}_{+}^{(n)}\subset\mathbb{Z}^{(n)}$
from the multi-object state $\boldsymbol{X}_{+}$ that evolved from
$\boldsymbol{X}$ according to the multi-object transition density
$\boldsymbol{f}_{+}(\boldsymbol{X}_{+}|\boldsymbol{X},a)$. The multi-agent
multi-object observation $\boldsymbol{\,Z}_{+}\in\pmb{\mathcal{Z}}$
is given by $\uplus_{n\in \mathcal{N}}\big({Z}_{+}^{(n)} \times \{ n\}\big)$, or equivalently the $|\mathcal{N}|$-tuple $({Z}_{+}^{(1)} \times \{1\},\dots,{Z}_{+}^{(|\mathcal{N}|)}\times \{|\mathcal{N}|\})$,
and has likelihood $\boldsymbol{g}_{+}(\boldsymbol{Z}_{+}|\boldsymbol{X}_{+},a)$. Note that, by convention ${Z}_{+}^{(n)} \times \{ n\} = \emptyset$ when agent $n$ is inactive (broken down), which is different from ${Z}_{+}^{(n)} \times \{ n\} = \emptyset \times \{n\}$ when $n$ is active but collects the empty observation (e.g., because there is nothing in its FoV). The specifics of $\boldsymbol{f}_{+}$ and $\boldsymbol{g}_{+}$ are
detailed in Subsection \eqref{subsec:Object-Dynamics}.

The \textit{information state} of the MPOMDP at time $j>1$ is the
(labelled) \textit{multi-object filtering density} $\boldsymbol{\pi}_{a_{j-1},j}(\cdot|\boldsymbol{Z}_{j})$,
which contains all information about the multi-object state given
the observation history \cite{vo2013glmb}. For notational convenience,
we indicate dependence only on the latest control action and observation.
Starting with the initial prior $\boldsymbol{\pi}_{a_{0},1}$, the
information state $\boldsymbol{\pi}_{a_{j-1},j}(\cdot|\boldsymbol{Z}_{j})$
is propagated via Bayes recursion: 
\begin{align}
&\boldsymbol{\pi}_{a_{j-1},j}(\boldsymbol{\boldsymbol{X}}_{j})  =\int\!\!\boldsymbol{f}_{j}(\boldsymbol{X}_{j}|\boldsymbol{X},a_{j-1})\boldsymbol{\pi}_{a_{j-2},j-1}\!\left(\boldsymbol{X}\right)\delta\boldsymbol{X},\label{e:MTBF1}\\
&\boldsymbol{\pi}_{a_{j-1},j}(\boldsymbol{\boldsymbol{X}}_{j}|\boldsymbol{Z}_{j})  =\frac{\boldsymbol{g}_{j}(\boldsymbol{Z}_{j}|\boldsymbol{\boldsymbol{X}}_{\!j},a_{j-1})\boldsymbol{\pi}_{a_{j-1},j}(\boldsymbol{\boldsymbol{X}}_{j})}{\boldsymbol{g}_{j}(\boldsymbol{Z}_{j}|a_{j-1})},\label{e:MTBF2}
\end{align}
where $\boldsymbol{g}_{j}(\boldsymbol{Z}_{j}|a_{j-1})=\int\boldsymbol{g}_{\!j}(\boldsymbol{Z}_{j}|\boldsymbol{\boldsymbol{X}}_{\!j},a_{j-1})\boldsymbol{\pi}_{a_{j-1},j}(\boldsymbol{\boldsymbol{X}}_{\!j})\delta\boldsymbol{\boldsymbol{X}}_{j}$
is the \textit{predictive likelihood} under action $a_{j-1}$, and
the integral is Mahler's set integral (see also Subsection \ref{subsec:Object-DiffEnt}
for details).

A POMDP seeks the control action sequence $a_{k:k+H-1}$ that maximises
a \textit{value function} $V\big(a_{k:k+H-1}\big)$ constructed from
the information states and immediate rewards over horizon $H$, e.g.,
the expected sum of immediate rewards 
\begin{alignat}{1}
 & V\big(a_{k:k+H-1}\big)=\label{eq:opt_act}\\
 & \sum_{j=k+1}^{k+H}\int\!\!\!\int\!\!\boldsymbol{\varrho}_{j}\!\left(\boldsymbol{Z},\boldsymbol{X}\!,a_{j-1}\right)\boldsymbol{\pi}_{\!a_{j-1},j}(\boldsymbol{X}|\boldsymbol{Z})\boldsymbol{g}_{j}(\boldsymbol{Z}|a_{j-1})\delta\boldsymbol{X}\delta\boldsymbol{\boldsymbol{Z}}\!.\nonumber 
\end{alignat}
In general, the value function \eqref{eq:opt_act} cannot be evaluated
analytically, while numerical evaluations are intractable \cite{hauskrecht2000value}. Especially for the general case, the control action $a_j$ can be changed at each time step. Consequently, the action space grows exponentially over the horizon,  leading to the intractability of computing the value function \eqref{eq:opt_act}. An approximate strategy is to assume $a_{j}=a$ for $j=k:k+H-1$, and determine the action $a$ that optimises the resulting value function, but only apply this action for $M<H$ time steps. The process is then repeated at time $M+1$. For single-angle planning for tracking problems, this approximate search strategy has shown to be effective in solving practical problems\cite{hoa2019jofr,chen2024conservationbots}.  However, for multi-object multi-agent problems, this approximation strategy is still too computationally intensive~\cite{mahler2019glmb}.

In this work, we consider a computationally
tractable value function based on the notion of \textit{Predicted
Ideal Measurement Set} (PIMS)~\cite{mahler2014advances}. Using the
information state $\boldsymbol{\pi}_{a_{k-1},k}(\cdot|\boldsymbol{Z}_{k})$
at time $k$, we first compute the multi-object prediction density
\begin{align*}
\widehat{\boldsymbol{\pi}}_{a,j}(\boldsymbol{\boldsymbol{X}}_{\!j}) & =\int\!\prod_{i=k+1}^{j}\!\!\boldsymbol{f}_{\!i}(\boldsymbol{X}_{\!i}|\boldsymbol{X}_{\!i-1},a)\boldsymbol{\pi}_{a_{k-1},k}\!\left(\boldsymbol{X}_{\!k}\right)\delta\boldsymbol{X}_{\!k:j-1},
\end{align*}
from which the \textit{predicted multi-object state} $\boldsymbol{\boldsymbol{\widehat{X}}}_{a,j}$
is determined via some multi-object estimator. The PIMS $\widehat{\boldsymbol{Z}}_{a,j}$
is the multi-object observation of $\boldsymbol{\boldsymbol{\widehat{X}}}_{a,j}$
without noise nor false positives/negatives and known
(object-to-measurement) origins, after action $a$ has been applied
at times $k$ to $k+j-1$ \cite{mahler2014advances}. The PIMS-based
value function is defined as 
\begin{align}
\!\!\!V(a)=\sum_{j=k+1}^{k+H}\int\!\boldsymbol{\varrho}_{j\!}\big(\widehat{\boldsymbol{Z}}_{a,j},\boldsymbol{\boldsymbol{X}}_{\!j},a\big)T\boldsymbol{(\pi}_{a,j}(\boldsymbol{X}_{\!j}|\widehat{\boldsymbol{Z}}_{a,j}))\delta\boldsymbol{\boldsymbol{X}}_{\!j},\!\!\label{eq:pims_def}
\end{align}
where $T$ is some transformation of the information state, e.g.,
a distribution constructed from its moments.  The
ideal measurements are the most informative measurements possible.
Real sensor observations (with noise, false positives/negatives and
unknown origins) ``close'' to the ideal measurements are more informative
than those further away. Hence optimising the PIMS
value function would yield actions that result in the agents collecting
informative observations.

\begin{remark}
The exclusion of noise and false positives/negatives
(present in real sensors) from the ideal measurements does not mean
the PIMS approach neglects the limitations of real sensors. Rather,
the rationale is to reward actions that result in (real) observations
with less noise and false positives/negatives. 
\end{remark}

\subsection{MS-GLMB Filtering \label{subsec:Object-Dynamics}}

\subsubsection{Multi-Object Dynamic Model}

The multi-object transition density $\boldsymbol{f}_{+}$ captures
the motions and births/deaths of objects. At time $k$, an object
with state $\boldsymbol{x}=(x,\ell)\in\boldsymbol{X}$ either survives
to the next time with probability $r_{S,+}(\boldsymbol{x})$ or dies
with probability $1-r_{S,+}(\boldsymbol{x})$ \cite{vo2013glmb}.
Conditional on survival it takes on the state $\boldsymbol{x}_{+}=(x_{+},\ell_{+})$
according to the transition density $f_{+}(x_{+}|x,\ell)\delta_{\ell}[\ell_{+}]$,
where $\delta_{\ell}[\ell_{+}]$ ensures the label remains the same.
In addition, an object with state $\boldsymbol{x}_{+}=(x_{+},\ell_{+})$
is born at time $k+1$ with probability $r_{B,+}(\ell_{+})$, and
its unlabelled state $x_{+}$ is distributed according to the probability
density $p_{B,+}(\cdot,\ell)$. It is standard practice to assume
that, conditional on the current multi-object state, objects are born
or displaced at the next time, independently of one another \cite{mahler2014advances}.
For the search and track problem considered here, the objects themselves
are not influenced by the multi-agent actions (i.e., the objects are not intelligent), and hence the multi-object
transition density $\boldsymbol{f}_{+}(\boldsymbol{X}_{+}|\boldsymbol{X})$
is independent of $a$. This also implies the independence of agent $n$'s observation 
from other agents' actions.
An explicit expression for $\boldsymbol{f}_{+}\!\left(\boldsymbol{X}_{+}|\boldsymbol{X}\right)$
based on the described model can be found in \cite{vo2013glmb}, though
this is not required in this work.

\subsubsection{Multi-Object Observation Model}

The multi-object observation likelihood $\boldsymbol{g}$ captures
detections/misdections, false alarms (or clutter), and data association
uncertainty. Given a multi-object state $\boldsymbol{X}$, each $\boldsymbol{x}\in\boldsymbol{X}$
has probability $P_{D}^{(n)}(\boldsymbol{x},u^{(n)})$ of being detected
by agent $n$ with state $u^{(n)}$, and generates an observation
$\boldsymbol{z}=(z,n')\in\pmb{\mathbb{Z}}$ with likelihood~\cite{mahler2019glmb}:
\begin{align*}
g^{(n)}(\boldsymbol{z}|\boldsymbol{x},u^{(n)})=\delta_{n}[n']g^{(n)}(z|\boldsymbol{x},u^{(n)}),
\end{align*}
or missed with probability $1-P_{D}^{(n)}(\boldsymbol{x},u^{(n)})$.
The term $\delta_{n}[n']$ ensures that detections recorded by agent
$n$ are tagged with label $n$. The observation set
collected by this agent is formed by the superposition of the (object-originated)
detections and Poisson clutter with intensity 
\begin{align*}
\kappa^{(n)}(\boldsymbol{z}|u^{(n)})=\delta_{n}[n']\kappa^{(n)}(z|u^{(n)}).
\end{align*}
It is standard practice to assume that conditional on $\boldsymbol{X}$,
detections are independent of each other and clutter~\cite{mahler2007statistical},
and that the observations obtained by individual agents are independent~\cite{thrun2005probabilistic}.
\begin{remark}~\label{rm:1} Since the multi-agent state is determined
by the control action $a=(a^{(1)},...,a^{(|\mathcal{N}|)})$, we indicate
this dependence by writing $P_{D}^{(n)}(\boldsymbol{x},a^{(n)})\triangleq P_{D}^{(n)}(\boldsymbol{x},u^{(n)})$,
$\kappa^{(n)}(\boldsymbol{z}|a^{(n)})\triangleq\kappa^{(n)}(\boldsymbol{z}|u^{(n)})$,
and $g^{(n)}(\boldsymbol{z}|\boldsymbol{x},a^{(n)})\triangleq g^{(n)}(\boldsymbol{z}|\boldsymbol{x},u^{(n)})$,
unless otherwise stated. \end{remark}

Due to the unknown data association, it is necessary to consider all
multi-agent association maps, i.e., combinations of measurement-to-object
origins. More concisely, a multi-agent \textit{association map }$\gamma$
is an\textit{\emph{ }}$|\mathcal{N}|$-tuple of \textit{positive 1-1
maps}\footnote{Maps in which no two distinct arguments are mapped to the same positive
value so that distinct objects cannot share the same measurement.} 
\begin{align*}
\gamma^{\left(n\right)}\!:\mathbb{L}\rightarrow\{-1{\textstyle :}|\boldsymbol{Z}^{\left(n\right)}|\},\: & n=1,...,|\mathcal{N}|,
\end{align*}
where $\gamma^{\left(1\right)}(\ell)=...=\gamma^{\left(|\mathcal{N}|\right)}(\ell)=-1$~if
label $\ell$ is dead/unborn, $\gamma^{\left(n\right)}(\ell)=0$ if
object $\ell$ is not detected by agent $n$, and $\gamma^{\left(n\right)}(\ell)>0$
if object $\ell$ generates observation $\boldsymbol{z}_{\gamma^{\left(n\right)}(\ell)}$
at agent $n$ \cite{vo2019multisensor}. The positive 1-1 property
ensures each observation from agent $n$ originates from at most one
object. The space of all such multi-agent associations is denoted
as $\Gamma$, and the set $\mathcal{L}(\gamma)\triangleq\{\ell\in\mathbb{L}:\gamma^{\left(1\right)}(\ell),...,\gamma^{\left(|\mathcal{N}|\right)}(\ell)\geq0\}$
is called the \textit{live labels} of $\gamma$. It is assumed that
conditional on the multi-object state, the measurements from individual
agents are independent (i.e., there is no interference among agents
in the observation process)~\cite{thrun2005probabilistic}. An explicit
expression for $\boldsymbol{g}\left(\boldsymbol{Z}|\boldsymbol{\boldsymbol{X}},a\right)$
based on the described model can be found in \cite{vo2019multisensor},
though this expression is not needed in this work.

\subsubsection{MS-GLMB Filter}

The Bayes recursion (\ref{e:MTBF1})-(\ref{e:MTBF2}) admits an analytical
information state (multi-object filtering density) in the form of
a GLMB \cite{vo2013glmb} 
\begin{equation}
\boldsymbol{\pi}\left(\boldsymbol{X}\right)=\sum_{I,\xi}w^{\left(I,\xi\right)}\delta_{I}[\mathcal{L}\left(\boldsymbol{X}\right)]\left[p^{(\xi)}\right]^{\boldsymbol{X}},\label{e:GLMB-1}
\end{equation}
where: $\mathcal{L}\left(\boldsymbol{X}\right)$ denotes the labels
of $\boldsymbol{X}$; each $I$ is a finite subset of $\mathbb{L}$;
each $\xi$ is a history $\gamma_{1:k}$ of multi-agent association
maps up to the current time; each $w^{\left(I,\xi\right)}$ is non-negative
such that $\sum_{I,\xi}w^{\left(I,\xi\right)}=1$; and each $p^{\left(\xi\right)}\left(\cdot,\ell\right)$
is a probability density on the attribute space $\mathbb{X}$. Note
that the traditional GLMB form involves the distinct label indicator
$\delta_{\left|\boldsymbol{X}\right|}\left[\left|\mathcal{L}\left(\boldsymbol{X}\right)\right|\right]$,
which is not needed here since the multi-object state space is the
partition matroid $\pmb{\mathcal{X}}$. For convenience, we abbreviate
the GLMB in (\ref{e:GLMB-1}) by the set of components, i.e., $\boldsymbol{\pi}\triangleq\left\{ \left(w^{\left(I,\xi\right)},p^{\left(\xi\right)}\right)\right\} $.

Under the standard multi-object system model \cite{mahler2007statistical}
described above, the Bayes recursion propagates a GLMB information
state $\boldsymbol{\pi}$ to the next GLMB information state 
\begin{equation}
\boldsymbol{\pi}_{a,+}=\Omega\left(\boldsymbol{\pi};\boldsymbol{Z}_{+},a\right),\label{e:GLMB_operator}
\end{equation}
where $\Omega$ denotes the \textit{MS-GLMB recursion operator} (which
depends on the multi-object model parameters $r_{B},p_{B}$, $r_{S}$,
$f_{+}$, $\kappa^{\left(n\right)}$, $P_{D}^{(n)}$, and $g^{\left(n\right)}$,
$n\in\mathcal{N}$, see Appendix \ref{sec:ms_glmb_appendix}
for the actual mathematical expression). Thus, starting with an initial
GLMB, all subsequent information states are GLMBs. In practice, the
number of GLMB components grows with time and truncation is needed
to curb this growth \cite{vo2019multisensor}.

\section{Path Planning for Multi-Object Tracking }

\label{sec:planning} This section presents our approach to multi-agent
planning, with limited FoV sensors, for searching and tracking an
unknown and time-varying number of objects, as conceptualised in Fig.~\ref{fig:overall_planning_framework}.
Subsection \ref{subsec:Object-DiffEnt} extends the notion of differential
entropy and mutual information to RFS. Subsection \ref{subsec:LMB-DiffEnt}
discusses labelled multi-Bernoulli as the labelled first moment of
labelled RFS and derives its differential entropy. Building on this,
Subsections \ref{subec:planning_r1} and \ref{subsec:Planning-r2}
formulate the tracking and discovery value functions. The proposed
multi-agent path planning algorithm is presented in Subsections \ref{subsec:planning_rmo}
and \ref{subsubsec:greedy_search}.

\subsection{Differential Entropy \label{subsec:Object-DiffEnt}}

Differential entropy quantifies the uncertainty of a random variable,
with lower values indicating lower uncertainty \cite[pp.6]{cover2012elements}.
Further, in probabilistic planning, differential entropy can be used
to assess the information gained from new observations~\cite{driess2019active}.
Consequently, an extension of this concept to RFS is needed to formulate
the information-based planning objectives for our
MPOMDP, and to derive tractable solutions.

To define meaningful differential entropy for RFS, we need to revisit
the notion of probability density. The probability density of an RFS
is taken with respect to (w.r.t.) the reference measure $\mu$ defined
for each (measurable) $\Tau\subseteq\mathcal{X}$ (the class of finite
subsets of $\mathbb{X}$) by 
\begin{align*}
\mu(\Tau)\triangleq\sum_{i=0}^{\infty}\dfrac{1}{i!K^{i}}\int1_{\Tau}(\{y_{1},\dots,y_{i}\})d(y_{1},\dots,y_{i}),
\end{align*}
where $K$ is the unit of hyper-volume on $\mathbb{X}$, $1_{\Tau}(\cdot)$
is the indicator function for $\Tau$, and by convention the integral
for $i=0$ is $1_{\Tau}(\emptyset)$~\cite{vo2005sequential}. The
role of $\mu$ is analogous to the Lebesgue measure on Euclidean space,
and the integral of a function $f:\mathcal{X}\rightarrow(-\infty,\infty)$
w.r.t. $\mu$ is given by 
\begin{align*}
\int\!f(Y)\mu(dY) & =\sum_{i=0}^{\infty}\dfrac{1}{i!K^{i}}\int\!f(\{y_{1},\dots,y_{i}\})d(y_{1},\dots,y_{i}).
\end{align*}
Note that $f$ and $\mu$ are both dimensionless/unitless. The probability
density $f_{X}\!:\mathcal{X}\rightarrow\![0,\infty)$ of an RFS $X$
satisfies $\mathrm{{\textstyle Pr}}(X\!\in\!\Tau)=\int\!1_{\Tau}(Y)f_{X}(Y)\mu(dY)$
for each $\Tau\subseteq\mathcal{X}$.

The integral above is equivalent to Mahler's set integral (see Proposition
1 in ~\cite{vo2005sequential}), defined for a function $\pi$ by~\cite{mahler2007statistical}
\begin{align*}
\int\!\pi(Y)\delta Y\triangleq\sum_{i=0}^{\infty}\dfrac{1}{i!}\int\!\pi(\{y_{1},\dots,y_{i}\})d(y_{1},\dots,y_{i}),
\end{align*}
in the sense that $\int K^{-|Y|}f(Y)\delta Y=\int f(Y)\mu(dY)$. This
equivalence means that Mahler's multi-object density of the RFS $X$
is given by $\pi_{X}(Y)\triangleq K^{-|Y|}f_{X}(Y)$.

Equipped with the above construct of density/integration, the notion
of differential entropy (and mutual information) naturally extends
to RFS as follows.

\begin{definition} The \textit{differential entropy} $h(X)$ of an
RFS $X$, with probability density $f_{X}$, is defined as 
\begin{align}
h(X) & =-\mathbb{E}_{X}\left[\ln f_{X}\right]=-\!\int\!\ln\big(f_{X}(Y)\big)f_{X}(Y)\mu(dY).\label{eq:gen_diff_entropy}
\end{align}
\end{definition} \begin{remark}\label{remark:MI}Differential entropy
can be extended to a sequence of RFS $X_{1},...,X_{m}$, with joint
probability density $f_{X_{1},...,X_{m}}$ as $h(X_{1},...,X_{m})=-\mathbb{E}_{X_{1},...,X_{m}}\left[\ln f_{X_{1},...,X_{m}}\right]$,
and to conditional differential entropy of $X$ on $Z$, with conditional
probability density $f_{X|Z}$ as 
\begin{align}
h(X|Z) & =-\mathbb{E}_{X,Z}\left[\ln f_{X|Z}\right].\label{eq:gen_diff_entropy-1}
\end{align}
This notion of differential entropy means that the mutual information
between the RFSs $X$ and $Z$ is 
\begin{align}
I(X;Z) & =h(X)-h(X|Z).\label{eq:gen_mutual_information}
\end{align}
\end{remark} \begin{remark}Using Mahler's set integral, differential
entropy can be written in terms of the multi-object density $\pi_{X}$
as 
\begin{align}
\!\!h(X) & =-\!\int\!\ln\big(K^{|Y|}\pi_{X}(Y)\big)\pi_{X}(Y)\delta Y.\label{eq:gen_diff_entropy_fisst-1}
\end{align}
\end{remark} A low differential entropy $h(X)$ translates to low
uncertainty in the RFS $X$ \cite[pp.6]{cover2012elements}. Moreover,
given knowledge of another RFS $Z$, high mutual information $I(X;Z)$
means that observing $Z$ would provide more information (or reduce
uncertainty) on $X$, because $I(X;Z)$ quantifies the \textquotedbl amount
of information\textquotedbl{} obtained about $X$ by observing $Z$
\cite[pp.6]{cover2012elements}. Hence, from a state estimation context,
it is prudent to minimise the differential entropy of $X$, or maximise
its mutual information with the observation $Z$. %

\subsection{Differential Entropy for Labelled Multi-Bernoulli \label{subsec:LMB-DiffEnt}}

While differential entropy for labelled RFSs is computationally intractable
in general, for the special case of \textit{labelled multi-Bernoulli}
(LMB), this can be computed analytically. An LMB, with parameters
$\{r^{(\ell)},p^{(\ell)}(\cdot)\}_{\ell\in\mathbb{L}}$, has multi-object
density of the form 
\begin{align}
\boldsymbol{\widehat{\pi}}(\boldsymbol{X}) & =r^{\mathcal{L}(\boldsymbol{X})}\tilde{r}^{\mathbb{L}\setminus\mathcal{L}(\boldsymbol{X})}p^{\boldsymbol{X}},\label{eq:LMB_RFS-1}
\end{align}
where $r^{(\ell)}$ is the existence probability of object $\ell$,
$\tilde{r}^{(\ell)}=1-r^{(\ell)}$, and $p(x,\ell)=p^{(\ell)}(x)$
is the probability density of its attribute conditional on existence.
Like the Poisson, an LMB is \textit{completely characterised} by the
\textit{first-moment density}, commonly known as the Probability Hypothesis
Density (PHD), given by $\boldsymbol{v}(x,\ell)=r^{(\ell)}p^{(\ell)}(x)$,
specifically, $\boldsymbol{\widehat{\pi}}(\boldsymbol{X})=\boldsymbol{\widehat{\pi}}(\emptyset)\left[\boldsymbol{v}/\tilde{r}\right]^{\boldsymbol{X}}$,
i.e., a multi-object exponential of the PHD (assuming $r^{(\ell)}<1$).

Analogous to the Poisson, the LMB that matches the PHD of a labelled
RFS is treated as its \textit{labelled first moment}, e.g., the labelled
first moment of the GLMB $\boldsymbol{\pi}\triangleq\left\{ \left(w^{\left(I,\xi\right)},p^{\left(\xi\right)}\right)\right\} $
is the LMB with \cite{reuter2014lmb} 
\begin{align}
r^{(\ell)}= & \sum_{I,\xi}1_{I}(\ell)w^{(I,\xi)},\label{eq:GLMBexistenceprob}\\
p^{(\ell)}(x)= & \sum_{I,\xi}\frac{1_{I}(\ell)w^{(I,\xi)}p^{(\xi)}(x,\ell)}{r^{(\ell)}}\label{eq:GLMBtrackdensity}
\end{align}
The key benefits of the LMB over the Poisson (unlabeled first moment)
are the trajectory information and a cardinality variance that does
not grow with the mean.

The following Proposition establishes an analytic expression for the
differential entropy of the LMB, which can be evaluated with $\mathcal{O}(|\mathbb{L}|)$
complexity, i.e., linear in the number of labels. The proof is given
in Appendix~\ref{subsec:Proof-of-Tracking}. \begin{proposition}
\label{prop:diff_entropy_lmb} The differential entropy of an LMB
$\boldsymbol{X}$, with parameter set $\{r^{(\ell)},p^{(\ell)}(\cdot)\}_{\ell\in\mathbb{L}}$
is 
\begin{gather}
h(\boldsymbol{X})=\!-\sum_{\ell\in\mathbb{L}}\!\left[r^{(\ell)\!}\ln r^{(\ell)\!}+\tilde{r}^{(\ell)\!}\ln\tilde{r}^{(\ell)\!}+r^{(\ell)\!}\langle p^{(\ell)\!},\ln(Kp^{(\ell)})\rangle\right]\!.\label{eq:gen_diff_entropy_lmb}
\end{gather}
\end{proposition} \begin{remark}~\label{remark:Bernoulli_RFS_entropy}
A special case of Proposition~\ref{prop:diff_entropy_lmb} is the
differential entropy of a Bernoulli RFS,~\ie,~an LMB with only
one component, parameterised by $(r,p)$: 
\begin{gather*}
h(X)=-\big[r\ln(r)+\tilde{r}\ln\tilde{r}+r\langle p,\ln(Kp)\rangle\big].
\end{gather*}
\end{remark}

\vspace{-0.55cm}
 
\subsection{Tracking Value Function}

\label{subec:planning_r1}

This subsection presents the information-based value
function for the tracking task. Choosing the action that minimises
the mutual information between the multi-object state
and the observation (resulting from the action) reduces uncertainty
on the multi-object state and, hence, improves tracking accuracy.
Note from \eqref{eq:gen_mutual_information} that
the mutual information between the RFSs $X$ and $Z$ is $I(X;Z)=h(X)-h(X|Z)$.
Since $h(X)$ is independent of the control action $a$, maximising
the mutual information $I(X;Z)$ is equivalent to minimising $h(X|Z)$,
the differential entropy of $X$ conditioned on $Z$. 

For computational tractability, we adopt the PIMS approach~\cite{mahler2007statistical}.
Specifically, we use a tracking value function $V_{1}(\cdot)$ of
the form \eqref{eq:pims_def}, with immediate reward, at time $j$,
given by $\boldsymbol{\varrho}_{j}\big(\widehat{\boldsymbol{Z}}_{a,j},\boldsymbol{X}_{j},a\big)=\ln\!\big(K^{|\boldsymbol{X}|}\widehat{\boldsymbol{\pi}}_{a,j}(\boldsymbol{X}_{j}|\widehat{\boldsymbol{Z}}_{a,j})\big)$,
and $T\boldsymbol{(\pi}_{a,j}(\boldsymbol{X}_{j}|\widehat{\boldsymbol{Z}}_{a,j}))=\widehat{\boldsymbol{\pi}}_{a,j}(\boldsymbol{X}_{j}|{\widehat{\boldsymbol{Z}}_{a,j}})$, 
where $\widehat{\boldsymbol{\pi}}_{a,j}(\cdot|\widehat{\boldsymbol{Z}}_{a,j})$
denotes the labelled first moment of the information state $\boldsymbol{\pi}_{a,j}(\cdot|\widehat{\boldsymbol{Z}}_{a,j})$,
i.e., $\widehat{\boldsymbol{\pi}}_{a,j}(\cdot|\widehat{\boldsymbol{Z}}_{a,j})$
is the LMB matching the first moment of $\boldsymbol{\pi}_{a,j}(\cdot|\widehat{\boldsymbol{Z}}_{a,j})$.
This results in 
\begin{align}
V_{1}(a)=-\sum_{j=k+1}^{k+H}h(\boldsymbol{X}_{j}|\widehat{\boldsymbol{Z}}_{a,j}),\label{eq_R1_Track}
\end{align}
i.e., the cumulative differential entropy of the multi-object state
given the PIMS, over the horizon. 

It is important to note that, in addition to being a meaningful tracking
value function, $V_{1}(\cdot)$ can be evaluated analytically and
efficiently. Since the information state is a GLMB (see Subsection
\ref{subsec:Object-Dynamics}), the labelled first moment is the LMB $\widehat{\boldsymbol{\pi}}_{a,j}(\boldsymbol{X}_{j}|{\widehat{\boldsymbol{Z}}_{a,j}})$ whose parameters are given by \eqref{eq:GLMBexistenceprob}-\eqref{eq:GLMBtrackdensity}. 
Moreover, the differential entropy of this LMB can be evaluated in closed form, as given in \eqref{eq:gen_diff_entropy_lmb}, with a 
complexity that is \textit{linear} in the number of labels, i.e., 
$\mathcal{O}(|\mathbb{L}|)$. This linear complexity is achieved because each LMB component can be updated individually as the ideal measurement has a known origin and no false positives/negatives. The complexity of this step is $\mathcal{O}(1)$. In addition, due to \textbf{Proposition 1}, the complexity of computing the differential entropy is $\mathcal{O}(|\mathbb{L}|)$. Hence, the total complexity of computing $V_1$ is $\mathcal{O}(|\mathbb{L}|)$. This is a significant improvement in complexity compared to employing other multi-object filters.

\begin{remark} Using the expected value function \eqref{eq:opt_act}
with 
\begin{align*}
\boldsymbol{\varrho}_{j}\!\left(\boldsymbol{Z},\boldsymbol{X}\!,a_{j-1}\right)= & \ln\!\big(K^{|\boldsymbol{X}|}\boldsymbol{\pi}_{\!a_{j-1},j}(\boldsymbol{X}|\boldsymbol{Z})\big)\\
 & -\ln\!\int\!K^{|\boldsymbol{X}|}\boldsymbol{\pi}_{\!a_{j-1},j}(\boldsymbol{X}|\boldsymbol{Y})\boldsymbol{g}_{j}(\boldsymbol{Z}|a_{j-1})\delta\boldsymbol{Y},
\end{align*}
yields the cumulative mutual information between the multi-object
state and its observations over the horizon. However, as discussed
earlier, this value function is intractable to evaluate. \end{remark}

\vspace{-0.0cm}

\subsection{Occupancy-based Discovery Value Function\label{subsec:Planning-r2}}

This subsection presents the entropy-based value function for the
discovery task via an occupancy grid. Intuitively, the knowledge of
unexplored regions can be incorporated into the planning to increase
the likelihood of discovering new objects. While objects of interest
outside the sensor's FoVs are undetected, this knowledge has not been
exploited for discovery. To this end, we develop a dynamic occupancy
grid to capture knowledge of undetected objects outside the sensor's
FoVs.

We partition the search area $G\subset\mathbb{R}^{d_{\varkappa}}$
into a grid~$\{\varkappa^{(i)}\}_{i=1}^{N}$, such that~ $\varkappa^{(i)}\cap\varkappa^{(j)}=\emptyset,i\neq j$,
and ~$G=\varkappa^{(1)}\cup\dots\cup\varkappa^{(N)}$. At time $k$,
the \textit{occupancy} of grid cell $\varkappa^{(i)}$ is modelled
as a Bernoulli random variable $O^{(i)}$, i.e., $O^{(i)}=1$ means
cell $\varkappa^{(i)}$ is occupied, and $O^{(i)}=0$ otherwise. Further,
let $\boldsymbol{Z}_{a_{-}}(\varkappa^{(i)})$ denotes the set of
multi-agent measurements originating from cell $\varkappa^{(i)}$
after action $a_{-}$ was taken at the previous time, and $Y_{a_{-}}^{(i)}=\delta_{\emptyset}[\boldsymbol{Z}_{a_{-}}(\varkappa^{(i)})]$
denotes the binary observation, which equals $1$
if no measurements originate from cell $\varkappa^{(i)}$, and $0$
otherwise. We define the discovery value function $V_{2}$, for the
PIMS approach, as the cumulative differential
entropy (which is also the Shannon entropy when the random variable
is discrete) of the occupancy grid over the horizon 
\begin{gather}
V_{2}(a)=-\sum_{j=k+1}^{k+H}\sum_{i=1}^{N}h(O_{j}^{(i)}|Y_{a,j}^{(i)}).\label{eq_R2_discover}
\end{gather}
Intuitively, for discovery, we are only interested in the occupancy
of cells in which objects are undetected, because
cells with detected objects have already been explored. The following
result shows that the entropy $h(O_{j}^{(i)}|Y_{a,j}^{(i)})$ only
depends on $\omega_{j}^{(i)}(a)\triangleq\textrm{Pr}(O_{j}^{(i)}=1|Y_{a,j}^{(i)}=1)$,
i.e., the probability that $\varkappa^{(i)}$ is occupied by undetected
objects. Moreover, the discovery value function \eqref{eq_R2_discover}
can be computed analytically by propagating an initial $\omega_{k}^{(i)}(a_{k-1})$
from $k$ to $j=k+1,...,k+H$, for each cell. The proof is given in
Appendix~\ref{subsec:Proof-of-Discovery}.

\begin{proposition}\label{prop:diff_entropy_occ} Let $P_{S,+}^{(i)}$
be the probability that at least one undetected object in $\varkappa^{(i)}$
is still there at the next time, $P_{B,+}^{(i)}$ ~be the probability
of at least one undetected object entering the cell at the next time,
and $Q_{D,+}^{(i)}(a)$ be the probability that objects in cell $\varkappa^{(i)}$
will not be detected by any agent at the next time if action $a$
is taken at the current time. Then, 
\begin{gather}
\!\!\!\!\!\!\!\!\!\!\!\!\!\!\!\!\!\!\!\!h(O_{j}^{(i)}|Y_{a,j}^{(i)})=-\left[1-\omega_{j-1}^{(i)}(a)+\omega_{j-1}^{(i)}(a)Q_{D,j}^{(i)}(a)\right]\nonumber \\
\!\!\!\!\!\!\!\!\!\!\quad\quad\times\!\left[(1-\omega_{j}^{(i)}(a))\!\ln(1-\omega_{j}^{(i)}(a))+\omega_{j}^{(i)}(a)\ln\omega_{j}^{(i)}(a)\right]\!.\label{eq:occupancy-entropy}
\end{gather}
where $Q_{D,j}^{(i)}(a)=\prod_{n\in\mathcal{N}}(1-P_{D,j}^{(i)}(a^{(n)}))$, and $P_{D,j}^{(i)}(a^{(n)})$ is the probability that agent $n$ detects
objects in cell $\varkappa^{(i)}$ at time $j$
when control action $a^{(n)}$ is taken. In other words, the probability of objects in cell $\varkappa^{(i)}$ are not detected by any agents depends on detection probabilities of cell $\varkappa^{(i)}$  from all agents for a given control action. Moreover, given $\omega_{j}^{(i)}(a_{j-1})$,
the next $\omega_{j+1}^{(i)}(a_{j})$ is given by 
\begin{eqnarray}
\!\!\!\!\!\!\!\!\!\omega_{j+1}^{(i)}(a_{j-1})\! & \!\!\!\!=\!\!\! & ((1-\omega_{j}^{(i)}(a_{j-1}))P_{B,+}^{(i)}+\omega_{j}^{(i)}(a_{j-1})P_{S,+}^{(i)},\label{eq:occ-pred}\\
\!\!\!\!\!\!\!\!\!\omega_{j+1}^{(i)}(a_{j})\! & \!\!\!\!=\!\!\!\!\! & \dfrac{\omega_{j+1}^{(i)}(a_{j-1})Q_{D,+}^{(i)}(a_{j})}{1-\omega{}_{j+1}^{(i)}(a_{j-1})+\omega_{j+1}^{(i)}(a_{j-1})Q_{D,+}^{(i)}(a_{j})}.\label{eq:occ-up}
\end{eqnarray}
\end{proposition}

Selecting the action(s) that minimises the differential entropy of the occupancy grid (i.e., maximise $V_{2}$) improves the likelihood of discovering previously undetected objects. Recently explored cells would have low entropy due to observations made by the agents. 
In contrast, the entropy of unexplored cells would be higher because of the lack of observations. Hence, actions that minimise the differential entropy of the occupancy grid tend to drive the agents to the unexplored cells.

\subsection{Multi-Objective Planning}

\label{subsec:planning_rmo} Multi-agent path planning with the competing
objectives of discovery and tracking can be naturally fulfilled by
multi-objective optimisation via the multi-objective value function
\begin{equation}
V(a)=[V_{1}(a),V_{2}(a)]^{\dagger},
\end{equation}
where $a\in\mathbb{A}$, $V_{1}$ and $V_{2}$ are respectively the
tracking and discovery value functions described in \eqref{eq_R1_Track}
and \eqref{eq_R2_discover}. Multi-objective optimisation identifies
meaningful trade-offs amongst objectives via the Pareto-set, wherein
no solutions can improve one objective without degrading the remaining
ones.

Since online planning requires finding a solution in a timely manner,
we adopt the global criterion method (GCM)~\cite{Koski1993}. This
is one of the standard computational techniques in multi-objective optimisation, which computes a trade-off solution based on the distance of the value functions from an ideal solution. In particular, GCM combines the two value functions into one by first normalising them, and then combining the normalised versions without weighting. Hence, GCM avoids pre-defining specific weights for each value function, which simplifies the decision-making process, leading to faster solutions. In scenarios without human intervention or preference information, GCM's equal weighting provides a neutral trade-off solution, avoiding bias towards any specific value function. Importantly, if the underlying value functions are submodular, GCM \textit{preserves} the submodularity property. 
The resulting value function for GCM, given by 
\begin{align}
V_{mo}(a)=\sum_{i=1}^{2}\dfrac{V_{i}(a)-\min\limits _{a\in\mathbb{A}}V_{i}(a)}{\max\limits _{a\in\mathbb{A}}V_{i}(a)-\min\limits _{a\in\mathbb{A}}V_{i}(a)},\label{eq_comb_reward}
\end{align}
yields a unique solution \cite{coello2007evolutionary}, and turns
the multi-objective optimisation problem into 
\begin{align}
\max_{a\in\mathbb{A}}V_{mo}(a).\label{eq_optimal_action}
\end{align}
In principle, solving problem \eqref{eq_optimal_action} is NP-hard
\cite{Messias2011}. Nonetheless, when $V_{mo}$ is \textit{monotone
submodular} (this holds when $V_{1}$ and $V_{2}$ are monotone submodular),
sub-optimal solutions with guaranteed optimality bound can be computed
using a greedy algorithm with drastically lower complexity~\cite{qu2019distributed,jawaid2015submodularity}.
An alternative to GCM is to seek a robust solution against the worst
possible objective \cite{krause2008robust}. However, this robust
submodular observation selection (RSOS) usually
results in a non-submodular value function. Another alternative is
a weighted sum of the value functions, which preserves submodularity,
but choosing a meaningful set of weights is an open problem.

\begin{algorithm}[!tb]
{\footnotesize{}{}{}{}{}{}{}{}\caption{Greedy Search}
\label{greedy_algo}}{\footnotesize\par}

{\footnotesize{}\begin{algorithmic}[1]  		 				
\State \textbf{Input}: $\pmb{\mathcal{A}}$, $V_{mo}$ \Comment{action space and value function.} 		 	
\State \textbf{Output}: $\boldsymbol{A}^g \in \pmb{\mathcal{A}} $ \Comment{greedy control action.} 			
\State $\boldsymbol{A}^g := \emptyset$ \Comment{initialise empty greedy control action.} 		 	
\State $P := \emptyset$ \Comment{initialise empty planned agent's list.} 		 				
\State $U := \mathcal{N}$ \Comment{initialise planning agent's list.} 		 				
\While{$U\not=\emptyset$}  		 				
\ForEach{ $n \in U$} 		 				
\State $[a_g^{(n)},V^{(n)}] := \argmax\limits_{a^{(n)} \in   \mathbb{A}^{(n)} }  V_{mo}(\boldsymbol{A}^g \uplus  a^{(n)})$  		 				
\EndFor 		 				
\State $n^* := \argmax\limits_{n \in U} V^{(n)}$ \Comment{$n^*$ that yields the best value function.}		
\State $\boldsymbol{A}^g := \boldsymbol{A}^g \uplus a^{(n^*)}_{g} $ \Comment{store action with agent $n^*$.} 		 				
\State $P := P \cup \{n^*\} $ \Comment{add $n^*$ into  planned agent's list.} 		 				
\State $U := U \setminus \{n^*\} $ \Comment{delete $n^*$ from  planning agents' list.} 		 	
\EndWhile 	 		
\end{algorithmic}}{\footnotesize\par}
\end{algorithm}

\subsection{Greedy Search Algorithm \label{subsubsec:greedy_search} }

Greedy search can provide suboptimal solutions to the MPOMDP in polynomial-time
with guaranteed optimality bound if the objective \textit{is monotone
submodular}~\cite{qu2019distributed,jawaid2015submodularity}. \begin{definition}
A set function $q$ is said to be \textit{monotone} if $\boldsymbol{A}\subseteq\boldsymbol{B}\Rightarrow q(\boldsymbol{A})\leq q(\boldsymbol{B})$,
and \textit{submodular} if $\boldsymbol{A}\subseteq\boldsymbol{B}$
 and $\boldsymbol{a}\notin\boldsymbol{B}$ $\Rightarrow q(\boldsymbol{B}\cup\{\boldsymbol{a}\})-q(\boldsymbol{B})\leq q(\boldsymbol{A}\cup\{\boldsymbol{a}\})-q(\boldsymbol{A})$
\cite{nemhauser1978analysis}. \end{definition} To show monotone
submodularity of the proposed value functions, we recast them as set
functions on the partition matroid $\pmb{\mathcal{A}}$ of the \emph{common
action space} $\pmb{\mathbb{A}}\triangleq\uplus_{n\in\mathcal{N}}\mathbb{A}^{(n)}\times\{n\}$.
For each $n\in\mathcal{N}$, recall that $V_{j}((a^{(1)},\dots,a^{(n)}))$,
$j=1,2,mo$, are the relevant value functions on the multi-agent action
space $\,\mathbb{A}^{(1)}\times...\times\mathbb{A}^{(n)}$, we define
corresponding value functions $\boldsymbol{V}_{j}(\cdot)$ on $\pmb{\mathcal{A}}$
by 
\begin{gather*}
\boldsymbol{V}_{j}(\{(a^{(m)},m)\}_{m=1}^{n})\triangleq V_{j}((a^{(1)},\dots,a^{(n)})).
\end{gather*}
Note that the subset $\,\pmb{\mathcal{A}}^{(n)}=\{\boldsymbol{A}\in\pmb{\mathcal{A}}:|\boldsymbol{A}|=n\}$
of $\pmb{\mathcal{A}}$ is equivalent to $\,\mathbb{A}^{(1)}\times...\times\mathbb{A}^{(n)}$
because any multi-agent action $(a^{(1)},\dots,a^{(n)})$ has a 1-1
correspondence with $\{(a^{(m)},m)\}_{m=1}^{n}\in\,\pmb{\mathcal{A}}^{(n)}$.
Thus, problem \eqref{eq_optimal_action} is equivalent to maximising
$\boldsymbol{V}_{mo}(\boldsymbol{A})$ over the partition matroid
$\pmb{\mathcal{A}}$ subject to the cardinality constraint $|\boldsymbol{A}|=|\mathcal{N}|$.
Here, using a set function facilitates the analysis 
of the greedy search algorithm (Algorithm~\ref{greedy_algo}) that
iteratively appends one single agent at each step (line $11$). 
\begin{proposition}
\label{prop:mutual_information_submodular} The mutual information
$I\big(\boldsymbol{X};\boldsymbol{Z}_{\boldsymbol{A}}\big)$ between
the multi-object state $\boldsymbol{X}$ and multi-agent measurement
$\boldsymbol{Z}_{\boldsymbol{A}}$ collected by the agents under action
$\boldsymbol{A}\in\pmb{\mathcal{A}}$, is a monotone submodular set
function on $\pmb{\mathcal{A}}$ (\emph{see Appendix \ref{subsec:Proof-of-Submodularity}
for proof}). 
\end{proposition} 

The above Proposition implies that
the conditional differential entropy $-h(\boldsymbol{X}|\boldsymbol{Z}_{\boldsymbol{A}})$
is also a monotone submodular set function, because $I\big(\boldsymbol{X};\boldsymbol{Z}_{\boldsymbol{A}}\big)=h(\boldsymbol{X})-h(\boldsymbol{X}|\boldsymbol{Z}_{\boldsymbol{A}})$
and $h(\boldsymbol{X})$ is independent of $\boldsymbol{A}$. Moreover,
since $\boldsymbol{V}_{1}(\boldsymbol{A})$ is a positive linear combination
of $-h(\boldsymbol{X}|\widehat{\boldsymbol{Z}}_{\boldsymbol{A}})$,
using \cite[pp.272]{nemhauser1978analysis}, we have the following
result. \begin{proposition}\label{coro:V1} The tracking value function
$\boldsymbol{V}_{1}$ is a monotone submodular set function. \end{proposition}

In addition, the same result also holds for the discovery value function
(\emph{see Appendix \ref{subsec:Proof-of-Submodularity} for proof}).
\begin{proposition}\label{theorem:V2} The discovery value function
$\boldsymbol{V}_{2}$ is a monotone submodular set function. \end{proposition}

It follows from the above propositions that the objective $\boldsymbol{V}_{mo}$
is also a monotone submodular function on $\pmb{\mathcal{A}}$, because
it is a positive linear combination of $\boldsymbol{V}_{1}$ and $\boldsymbol{V}_{2}$
\cite[pp.272]{nemhauser1978analysis}. This means that the inexpensive
Greedy Search (see \textbf{Algorithm
\ref{greedy_algo}}) can be used to compute a suboptimal solution, with the following optimality bound \cite{fisher1978analysis}.
\begin{proposition}\label{theorem_lower_bound} Let $\mathring{\boldsymbol{V}}_{mo}=\max_{\boldsymbol{A}\in\pmb{\mathcal{A}},|\boldsymbol{A}|=|\mathcal{N}|}\boldsymbol{V}_{mo}(\boldsymbol{A})$,
and $\boldsymbol{A}^{*}$ denote a solution computed via the Greedy
Search Algorithm \ref{greedy_algo}. Then 
\begin{align}
(1-e^{-1})\mathring{\boldsymbol{V}}_{mo}\leq\boldsymbol{V}_{mo}(\boldsymbol{A}^{*})\leq\mathring{\boldsymbol{V}}_{mo}.\label{eq_theorem_lower_bound}
\end{align}
\end{proposition} \begin{remark} Computing the optimal multi-agent
control action via exhaustive search incurs an $\mathcal{O}(|\mathbb{A}|^{|\mathcal{N}|}|\mathcal{N}|(|\mathbb{L}|+N)H)$
complexity. The above Proposition enables suboptimal solutions to
be computed via greedy search with a tight optimality bound, but at
a reduced $\mathcal{O}\left(|\mathbb{A}||\mathcal{N}|^{2}(|\mathbb{L}|+N)H\right)$
complexity. This is a drastic reduction from exponential complexity
to linear in the number of control actions, and quadratic in the number
of agents. \end{remark}

\section{Performance Evaluations }

\label{sec:real_dataset}This section demonstrates
the performance of our proposed multi-agent solution via numerical
experiments. Subsections \ref{subsec:CRAWDAD-Dataset-Experiment} and
\ref{subsec:Results-and-Discussions} present the results and discussions
of simulated experiments on the CRAWDAD dataset of taxi trajectories
in 
the San Francisco
Bay Area~\cite{piorkowski2009crawdad}. Further tests on simple yet
challenging hypothetical scenarios are presented in Subsection \ref{subsec:Challenging-Hypothetical-Scenari}. Simulations were run on a workstation with two AMD EPYC 7702 processors (each with 64 cores @ 2.0 GHz), 1 TB of RAM, and MATLAB  2022a.

\begin{figure}[!tb]
\centering \includegraphics[width=0.47\textwidth]{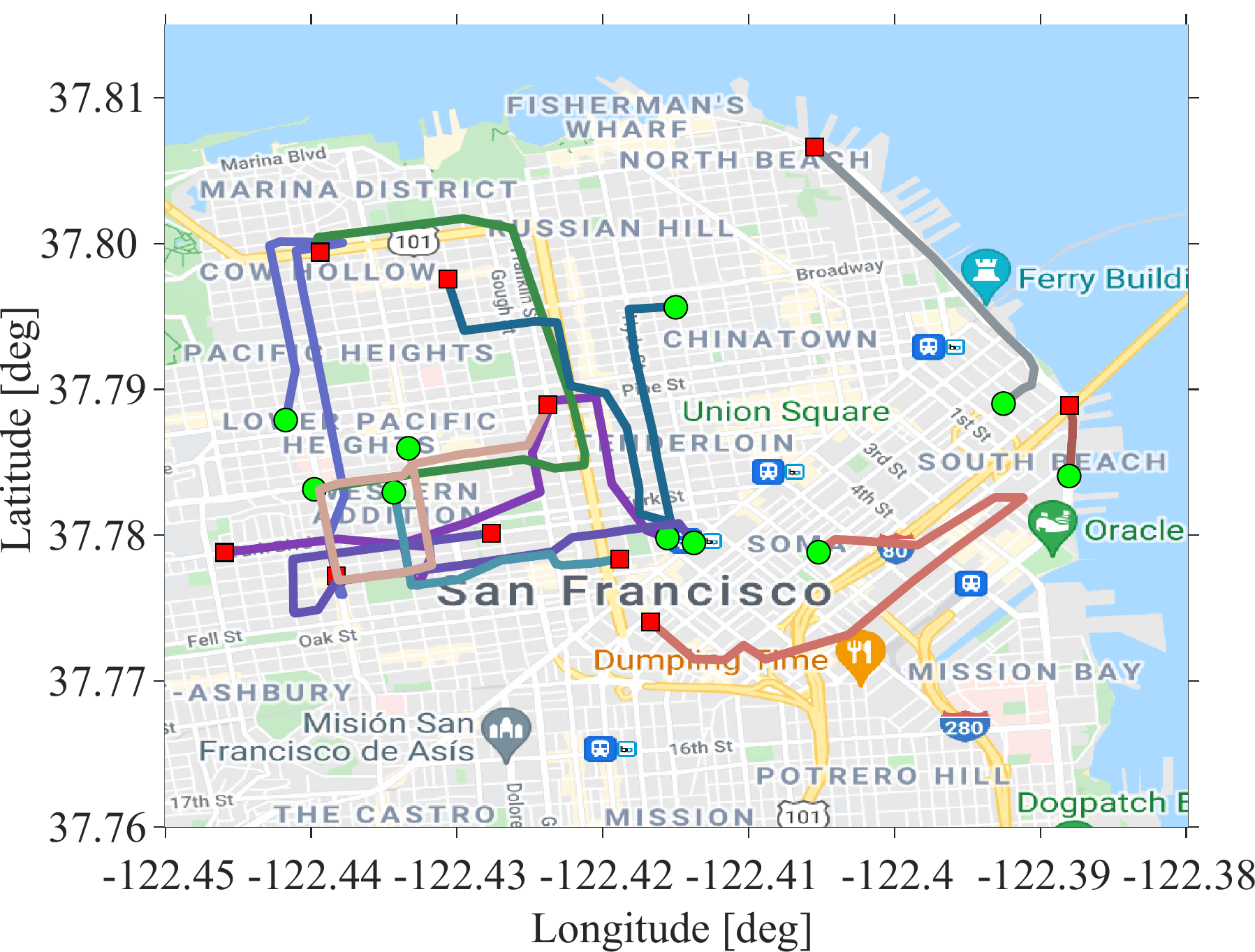}
\vspace{-0.1cm}
 \caption{CRAWDAD taxi dataset: $10$ taxis travelling over a $1000$~s period,
within an area of $6150~\text{m}\times6080$~m. The symbols $\bigcirc/\Box$
indicate the Start/Stop positions of each taxi, the different colours
representing different taxis}
. \label{fig:senario5_truth_gps}

\vspace{-0.3cm}
 
\end{figure}

\subsection{CRAWDAD Dataset Experiment Settings \label{subsec:CRAWDAD-Dataset-Experiment}}

The CRAWDAD dataset's real-world taxi trajectories,
coupled with synthetic measurements, allow us to control various
experimental parameters, especially with the time-varying number of
agents and objects. In particular, we randomly selected $10$ taxi
tracks over $1000$~s (from 18-May-2008 4:43:20
PM to 18-May-2008 5:00:00 PM), as shown in Fig.~\ref{fig:senario5_truth_gps}.
Following~\cite{dames2017detecting}, the $6150~\text{m}\times6080~\text{m}$ search area is scaled down
by a factor of $5$, while the time is sped up by 5 so that the taxi's
speed remains the same as in the real world. Thus, the simulated environment
has an area of $1230~\text{m}\times1216$~m and a total search time
of $200$~s. 

Since taxis can turn into different streets, we employ a constant
turn (CT) model with an unknown turn rate to account for this. In
particular, let $\boldsymbol{x}=(x,\ell)$ denote the single object
state comprising the label $\ell$ and kinematics $x=[\tilde{x}^{\dagger},\theta]^{\dagger}$
where $\tilde{x}=[\rho_{x},\dot{\rho}_{x},\rho_{y},\dot{\rho}_{y}]^{\dagger}$
is its position and velocity in Cartesian coordinates, and $\theta$
is the turn rate. Each object follows the constant turn model given
by 
\begin{align}
\tilde{x}_{k+1|k} & =F^{CT}(\theta_{k})\tilde{x}_{k}+G^{CT}\eta_{k},\\
\theta_{k+1|k} & =\theta_{k}+T_{0}q_{k}
\end{align}
where 
\begin{align}
F^{CT}(\theta)= & \begin{bmatrix}1 & \dfrac{\sin(\theta T_{0})}{\theta} & 0 & -\dfrac{1-\cos(\theta T_{0})}{\theta}\\
0 & \cos(\theta T_{0}) & 0 & -\sin(\theta T_{0})\\
0 & \dfrac{1-\cos(\theta T_{0})}{\theta} & 1 & \dfrac{\sin(\theta T_{0})}{\theta}\\
0 & \sin(\theta T_{0}) & 0 & \cos(\theta T_{0}),
\end{bmatrix},\\
G^{CT}= & \begin{bmatrix}T_{0}^{2}/2 & T_{0} & 0 & 0\\
0 & 0 & T_{0}^{2}/2 & T_{0}
\end{bmatrix}^{\dagger},
\end{align}

\noindent $T_{0}=1$~s is the sampling interval, $\eta_{k}\sim\mathcal{G}(\cdot;0,0.15^{2}I_{2})$
is Gaussian noise with $I_{2}$ denoting the $2\times2$ identity
matrix, and $q_{k}\sim\mathcal{G}(\cdot;0,(\pi/60)^{2})$.

The sensor on each agent $n$ is range-limited to $r_{D}$ and its
detection probability follows: 
\begin{gather}
P_{D}^{(n)}(\boldsymbol{x},u^{(n)})=\begin{cases}
P_{D}^{\max} & d(\boldsymbol{x},u^{(n)})\leq r_{D},\\
\max(0,P_{D}^{\max}-(d(\boldsymbol{x},u^{(n)})-r_{D})\hbar) & \text{otherwise,}
\end{cases}\label{eq:pd_by_distance}
\end{gather}
where $d(\boldsymbol{x},u^{(n)})$ is the distance between the object
$\boldsymbol{x}$ and agent $u^{(n)}$. A \textit{position} sensor
is considered in our experiments wherein a detected object $\boldsymbol{x}$
yields a noisy position measurement
$z=\big[\rho_{x},\rho_{y}]^{\dagger}+v$ , with $v\sim\mathcal{G}(\cdot;0,R)$ and $R=\mathrm{diag}([\sigma_{x}^{2},\sigma_{y}^{2}])$.
The parameters for our experiments are selected according to real-world
settings in~\cite{guido2016evaluating}. The detection
range $r_{D}=126$~m and $\hbar=0.008$, while the false-alarm per scan $\lambda_c$ is
$0.0223$, and the maximum detection probability $P_{D}^{\max}=0.8825$
(see Table~VIII in~\cite{zhu2018urban}). The minimum altitude of the UAVs is $126~$m, and to ensure no collisions, each operates at a different height with a $5$~m vertical separation. We observe that
the estimation error from a UAV flying at $60$~m altitude using
a standard visual sensor is around $0.55$~m~\cite{guido2016evaluating}.
Since our UAVs fly at higher, and the measurement noise is proportional
to the UAV's altitude, we set the measurement noise $\sigma_{x}=\sigma_{y}=0.55\times126/60=1.115$~m. 

A maximum number of $10$ UAVs (\eg, quad-copters) is considered
in this experiment. These UAVs depart from the centre of the search
area,~\ie, $[615,608,126]^{\dagger}$. The control
action space $\mathbb{A}^{(n)}$ is $\{\leftarrow,\nwarrow,\uparrow,\nearrow,\rightarrow,\searrow,\downarrow,\swarrow,\circ\}, ~\forall n\in\mathcal{N}$,
which represents the prescribed headings of the drone while moving with a maximum speed of $20$~m/s or staying stationary
($\circ$) at the current position. Since there is no prior knowledge
of each object's state, we model initial births at time $k=0$ by
an LMB density with parameters $\{(r_{B,0}^{(i)},p_{B}^{(i)})\}_{i=1}^{N_{B}}$,
where $N_{B}=100$ is the number
of possible new births, $p_{B}^{(i)}=\mathcal{G}(\cdot;m_{B}^{(i)},\Sigma_{B})$,
with $m_{B}^{(i)}=[m_{B,x}^{(i)},0,m_{B,y}^{(i)},0]^{\dagger}$, and
$\Sigma_{B}=\mathrm{diag}([\Delta_{x}/2,1,\Delta_{y}/2,1])^{2}$.
In this experiment we use $m_{B,x}^{(i)}=\Delta_{x}/2$$+((i-1)$mod
$10)\Delta_{x}$, $m_{B,y}^{(i)}=\Delta_{y}/2$$+$$\left\lfloor i/10\right\rfloor \Delta_{y}$,
$\Delta_{x}=123$ m, and $\Delta_{y}=121.6$ m, with ``mod" denoting
the modulo operator and $\left\lfloor \cdot\right\rfloor $ denoting
the floor operator. For the next time step, we
use an adaptive birth procedure that incorporates the current grid
occupancy information at time $k$ into the birth probability at the
next time $k+1$. Note that, since the number of
occupancy grid cells $N$ can be significantly large ($10,000$
in this case), which increases the filtering time if there are
too many birth components, we propose resizing the grid resolution
from $N$ cells with occupancy probability $\{\omega^{(i)}(a)\}_{i=1}^{N}$
to $N_{B}$ cells with occupancy probability $\{\bar{\omega}^{(i)}\}_{i=1}^{N_{B}}$
where $N_{B}\ll N$, using bicubic interpolation (e.g., with the \textsf{imresize}
command in MATLAB) to efficiently improve the filtering time. The
birth existence probability is then updated by: 
\begin{align}
r_{B,+}^{(i)}=\dfrac{\big[1+\bar{\omega}^{(i)}/\max(\bar{\omega}^{(i)})\big]\big(\sum_{i=1}^{N_{B}}r_{B,0}^{(i)}\big)}{\sum_{i=1}^{N_{B}}\big[1+\bar{\omega}^{(i)}/\max(\bar{\omega}^{(i)})\big]},\label{eq:adaptive_birth}
\end{align}
to ensure that the number of births remains stable over time, i.e.,
maintaining a constant expected number of births $\sum_{i=1}^{N_{B}}r_{B,+}^{(i)}$.

To evaluate tracking performance, we use the optimal sub-pattern assignment
(OSPA\textsuperscript{(2)}) metric from \cite{beard2020a} with a
cut-off value of $c=50$~m and order $p=1$ over a window spanning
the entire experimental duration. All results are averaged over $100$
Monte Carlo runs. A smaller \textbf{OSPA}\textsuperscript{(2)}\textbf{
Dist (m)} indicates a better tracking performance, covering localisation
accuracy, cardinality, track fragmentation, and track switching errors.

\begin{figure}[!tb]
\centering \includegraphics[width=0.4\textwidth]{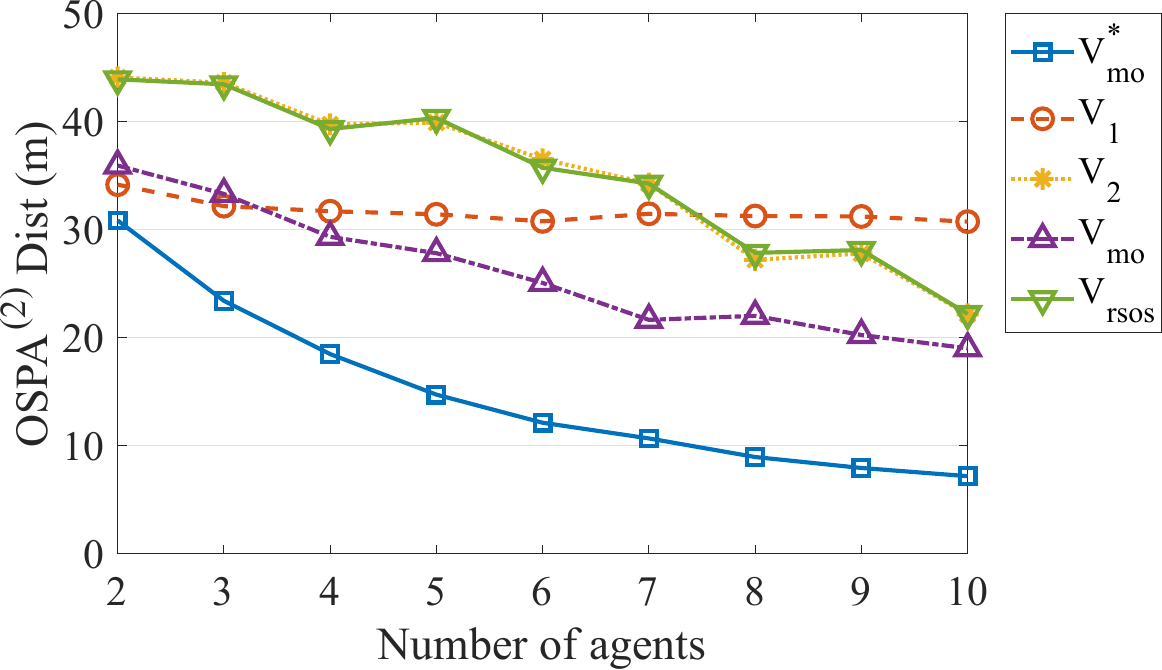}
\vspace{-0.2cm}
 \caption{Tracking performance of multi-objective planning
with $V_{mo}$ versus baseline methods in $100$ Monte Carlo runs
as $|\mathcal{N}|$ increases from $2$ to $10$. $V_{mo}^{*}$ represents
the optimal performance under ideal conditions where measurement origins
are known.}
\label{scenario5_ospa_dist_vs_ns} \vspace{-0.3cm}
 
\end{figure}

\begin{figure}[!tb]
\centering \includegraphics[width=0.4\textwidth]{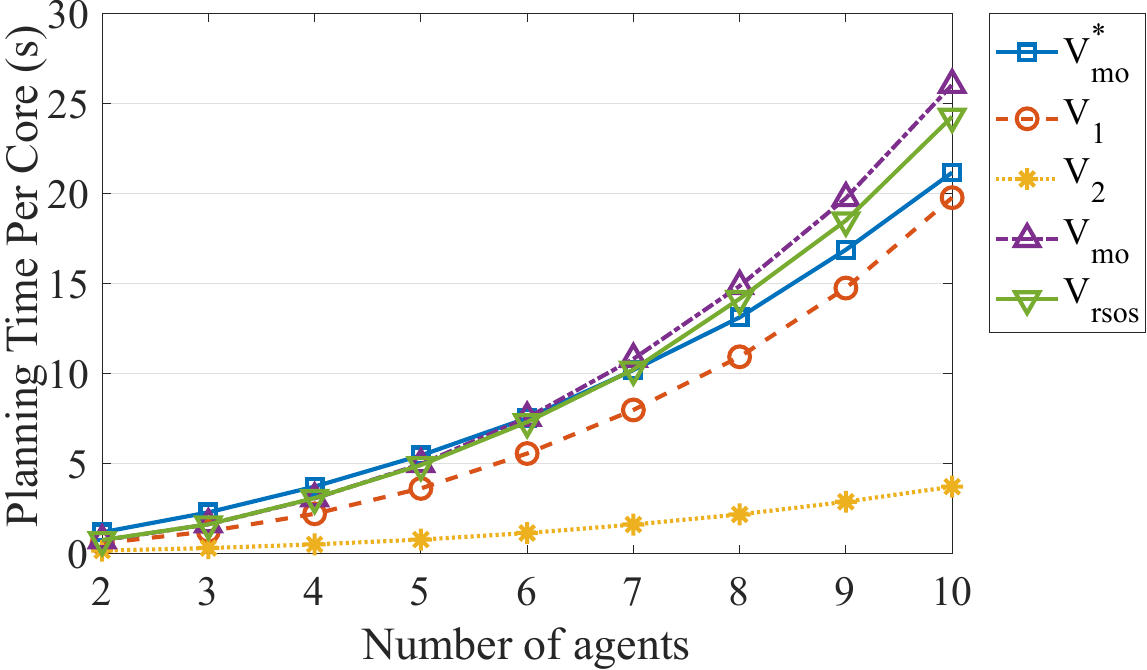}
\vspace{-0.3cm}
 \caption{Average planning time of multi-objective planning
with $V_{mo}$ versus baseline methods in $100$ Monte Carlo runs
as $|\mathcal{N}|$ increases from $2$ to $10$.}
\label{scenario5_plantime_vs_ns} \vspace{-0.3cm}
 
\end{figure}

\begin{figure}[!tb]
\vspace{-0.1cm}
\centering \includegraphics[width=1\columnwidth]{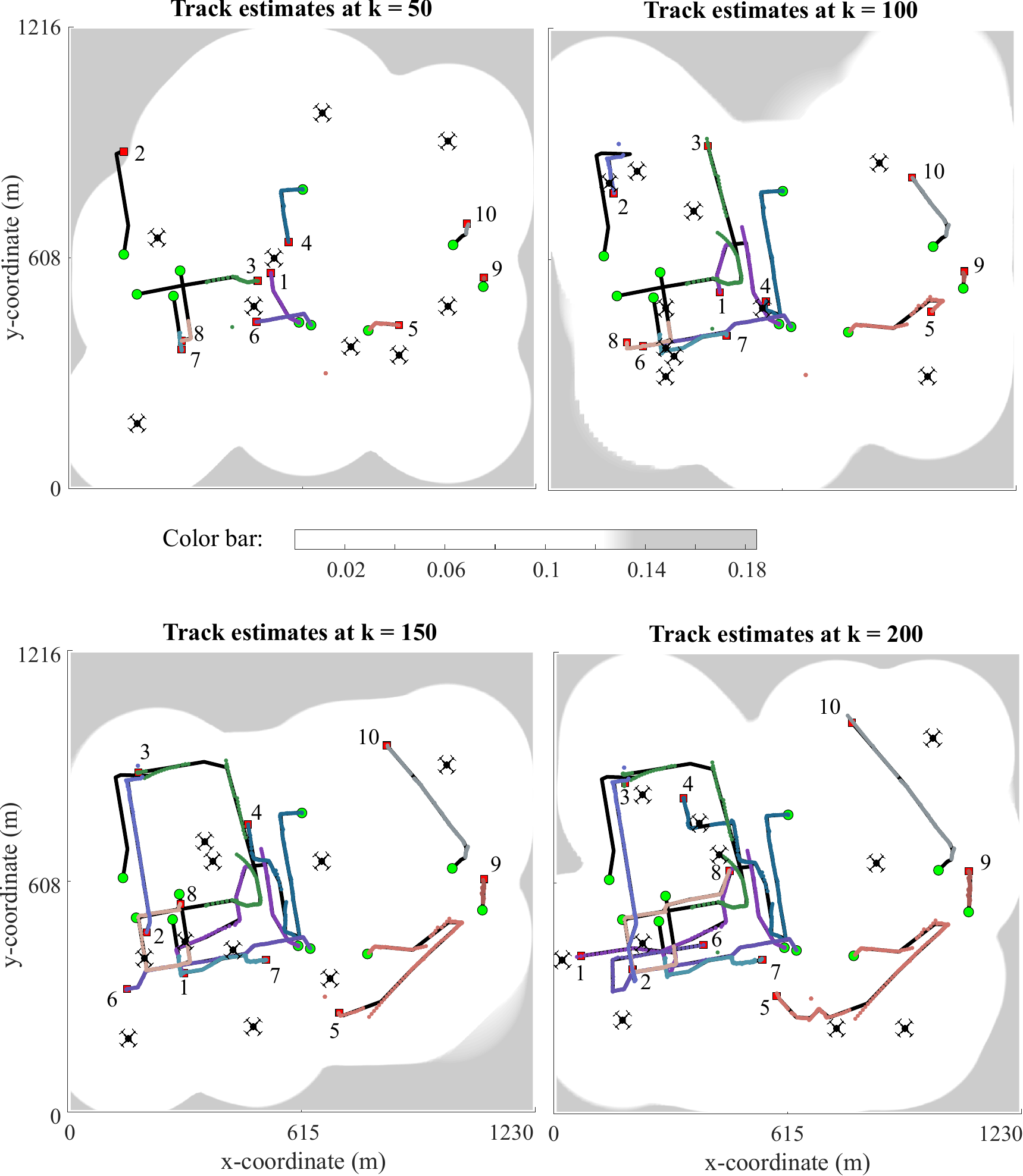}
\vspace{-0.6cm}
 \caption{A sample run for multi-objective planning with $V_{mo}$. \textbf{Background}:
grid occupancy probability. \textbf{Foreground}: estimated and true
positions of $10$ taxis. Black lines show the ground truth trajectories,
while the coloured dots denote the estimated positions. Circles ($\bigcirc$)
and squares ($\Box$) mark taxi start/stop points, with different
colours representing different taxi identities. The drone symbol indicates
the positions of the 10 UAVs.}
\label{fig:scenario5_rmo_particular_steps} \vspace{-0.3cm}
 
\end{figure}

\begin{figure}[!tb]
\centering \includegraphics[width=0.35\textwidth]{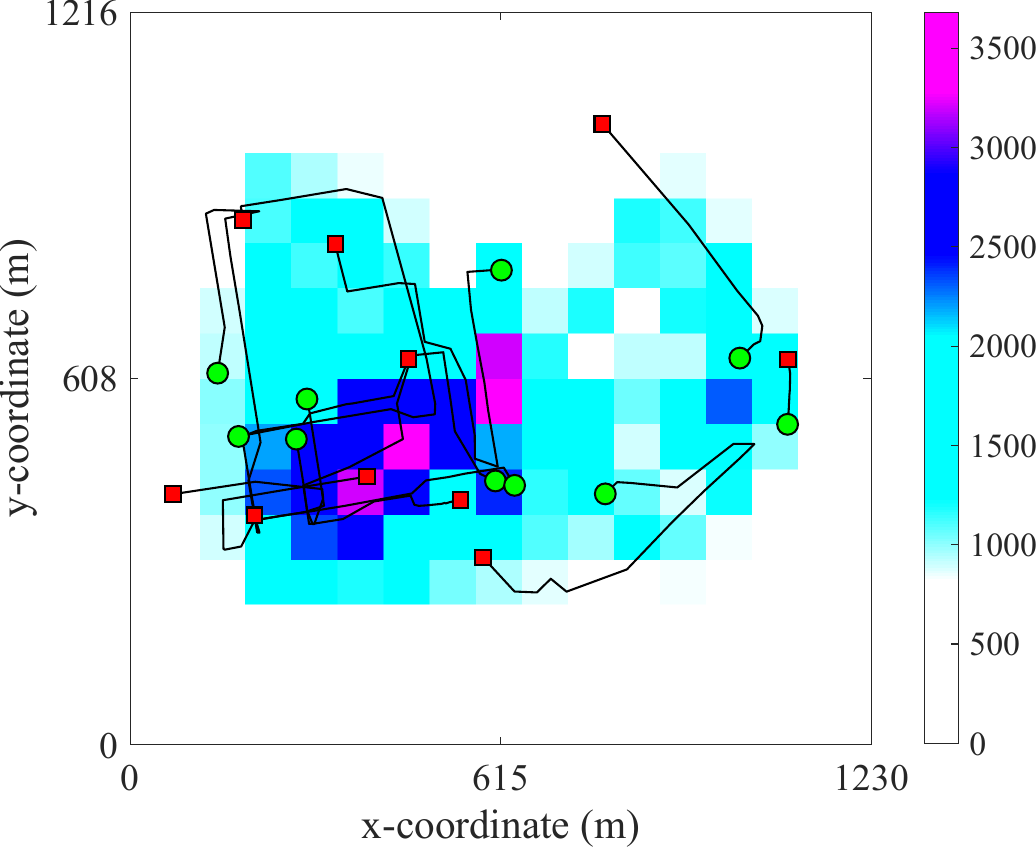}
\vspace{-0.3cm}
 \caption{Heatmap ($16\times16$ grids) showing the positions of $10$ UAVs
and taxi trajectories in multi-objective planning with $V_{mo}$,
over $100$ MC runs. Circles ($\bigcirc$) and squares ($\Box$) indicate
taxi start/stop positions. The colour legend represents UAV position
frequency.}
\label{fig:scenario5_heatmap} \vspace{-0.3cm}
 
\end{figure}
\vspace{-0.1cm}

\subsection{Results and Discussions\label{subsec:Results-and-Discussions}}

We first examine how the performance varies when the number of agents
increases, specifically, for the following three planning strategies:
\textbf{\textit{{(i)}}} \textit{{tracking}} only objective function
$V_{1}$ (conventional approach); \textbf{\textit{{(ii)}}} \textit{{discovering}}
only objective function $V_{2}$ (a special case of our approach);
and \textbf{\textit{{(iii)}}} multi-objective value function $V_{mo}$
that optimises trade-offs between tracking and discovering tasks (our
approach). Note that, the value function $V_{1}$ is a conventional
information-based method used in several previous works~\cite{dames2017detecting,binney2010informative,cliff2015online,hoffmann2009mobile,macdonald2019active,paull2012sensor},
and hence, considered as a baseline for comparisons. Additionally,
we benchmark the planning algorithms against the ideal case when data
association is known~\cite{hoa2020aaai}, i.e., the best possible
tracking performance. Furthermore, we compare $V_{mo}$
versus $V_{rsos}=\min(V_{1},V_{2})$, i.e., the robust submodular
observation selection (RSOS) method~\cite{krause2008robust} (ironically the RSOS method yields a \textit{non-submodular} value function).

Fig.~\ref{scenario5_ospa_dist_vs_ns} shows the \textbf{OSPA}\textsuperscript{(2)}
tracking error, over 100 Monte Carlo (MC) runs, for $10$ taxis in
the CRAWDAD taxi dataset when the number of agents is increased from
$2$ to $10$. On the one hand, when the number of agents is large
(more than four), our proposed (multi-objective planning with) $V_{mo}$
constantly outperforms (single-objective planning with) $V_{1}$ and
$V_{2}$, since there are enough agents to perform both tracking and
discovering tasks simultaneously. On the other hand, when the number
of agents is small (less than four), $V_{mo}$ achieves similar accuracy
as $V_{1}$ since there are not enough agents to cover the area, the
multi-agent focused on tracking instead of exploring. As expected,
$V_{1}$ does not improve tracking performance when the number of
agents increases since it only focuses on tracking detected objects
and misses objects outside of the team's FoVs. Overall, multi-objective
planning with $V_{mo}$ outperforms single-objective planning that
either focuses on tracking, i.e., $V_{1}$, or discovering, i.e.,
$V_{2}$. We also observe that $V_{mo}$ significantly
outperforms $V_{rsos}$. Moreover, since $V_{rsos}$ is not sub-modular, there is no performance bound for optimising $V_{rsos}$ using the greedy algorithm. As a result, we will \textit{exclude} $V_{rsos}$ from other experimental results.

\begin{table}[!tb]
\centering \caption{Performance comparison for different detection probabilities
and clutter rates when $|\mathcal{N}|=10$} 
\vspace{-0.3cm}
\label{tab:diff_pd_lambda} 
\begin{tabular}{|c|c||llll|}
\hline 
 &  & \multicolumn{4}{c|}{\textbf{OSPA}\textsuperscript{(2)}\textbf{Dist [m]}}\tabularnewline
\hline 
$P_{D}^{\max}$  & $\lambda_{c}$  & \multicolumn{1}{l|}{\textbf{$V_{1}$}} & \multicolumn{1}{l|}{\textbf{$V_{2}$}} & \multicolumn{1}{l|}{\textbf{$V_{mo}$}} & \textbf{$V_{mo}^{*}$}\tabularnewline
\hline 
\multirow{3}{*}{0.9} & 10  & \multicolumn{1}{l}{26.7} & \multicolumn{1}{l}{23.6} & \multicolumn{1}{l|}{\textbf{18.1}} & 10.1 \tabularnewline
\cline{2-6} \cline{3-6} \cline{4-6} \cline{5-6} \cline{6-6} 
 & 20  & \multicolumn{1}{l}{28.7} & \multicolumn{1}{l}{25.6} & \multicolumn{1}{l|}{\textbf{19.6}} & 11.3 \tabularnewline
\cline{2-6} \cline{3-6} \cline{4-6} \cline{5-6} \cline{6-6} 
 & 30  & \multicolumn{1}{l}{29.5} & \multicolumn{1}{l}{27.3} & \multicolumn{1}{l|}{\textbf{21.9}} & 12.0 \tabularnewline
\hline 
\multirow{3}{*}{0.8} & 10  & \multicolumn{1}{l}{29.3} & \multicolumn{1}{l}{29.9} & \multicolumn{1}{l|}{\textbf{20.0}} & 12.4 \tabularnewline
\cline{2-6} \cline{3-6} \cline{4-6} \cline{5-6} \cline{6-6} 
 & 20  & \multicolumn{1}{l}{31.8} & \multicolumn{1}{l}{31.3} & \multicolumn{1}{l|}{\textbf{22.1}} & 13.8 \tabularnewline
\cline{2-6} \cline{3-6} \cline{4-6} \cline{5-6} \cline{6-6} 
 & 30  & \multicolumn{1}{l}{32.3} & \multicolumn{1}{l}{31.8} & \multicolumn{1}{l|}{\textbf{24.5}} & 15.1 \tabularnewline
\hline 
\multirow{3}{*}{0.7} & 10  & \multicolumn{1}{l}{31.7} & \multicolumn{1}{l}{31.3} & \multicolumn{1}{l|}{\textbf{23.4}} & 15.4 \tabularnewline
\cline{2-6} \cline{3-6} \cline{4-6} \cline{5-6} \cline{6-6} 
 & 20  & \multicolumn{1}{l}{33.8} & \multicolumn{1}{l}{31.4} & \multicolumn{1}{l|}{\textbf{24.7}} & 17.4 \tabularnewline
\cline{2-6} \cline{3-6} \cline{4-6} \cline{5-6} \cline{6-6} 
 & 30  & \multicolumn{1}{l}{35.6} & \multicolumn{1}{l}{33.2} & \multicolumn{1}{l|}{\textbf{27.5}} & 18.8 \tabularnewline
\hline 
\end{tabular} 
\end{table}

Fig.~\ref{scenario5_plantime_vs_ns} presents the
averaged planning time using a single-core over 100 MC runs for different value
functions as the number of agents increases from $2$ to $10$. As
expected, $V_{2}$ exhibits the fastest planning time because it solely
focuses on discovering new objects via the occupancy grid. In contrast,
$V_{mo}$ incurs the longest planning time as it requires the computation of both $V_{1}$ and $V_{2}$. The result further supports the efficacy
of our proposed method, with planning time increasing proportionately
to the square of the number of agents. 

Table~\ref{tab:diff_pd_lambda} presents the tracking
performance for various detection probabilities and clutter rates
when the number of agents is fixed at 10, across 100 Monte Carlo (MC)
simulations. The results substantiate the robustness of our proposed
method ($V_{mo}$), which performs consistently better across various detection probabilities and clutter rates.
Further results for the 10-agent case demonstrate the effectiveness
of multi-objective planning. Fig.~\ref{fig:scenario5_rmo_particular_steps}
depicts the estimated versus true trajectories of the $10$ taxis
(from the CRAWDAD taxi dataset), and the occupancy probability, at
various times, for a particular run of the proposed (multi-objective
planning with) $V_{mo}$. The results demonstrate the correct search
and tracking of all taxis. The heat map over $100$ MC runs in Fig.~\ref{fig:scenario5_heatmap},
indicates that the $10$ agents concentrate on the western region
of the search area where there are more taxis around, while also slightly
covering the eastern region to successfully track the remaining taxis.
It is expected that as time increases, the agents start to spread
out from the centre to cover more areas for better discovery and tracking.

\begin{figure}[!tb]
\centering \includegraphics[width=0.35\textwidth]{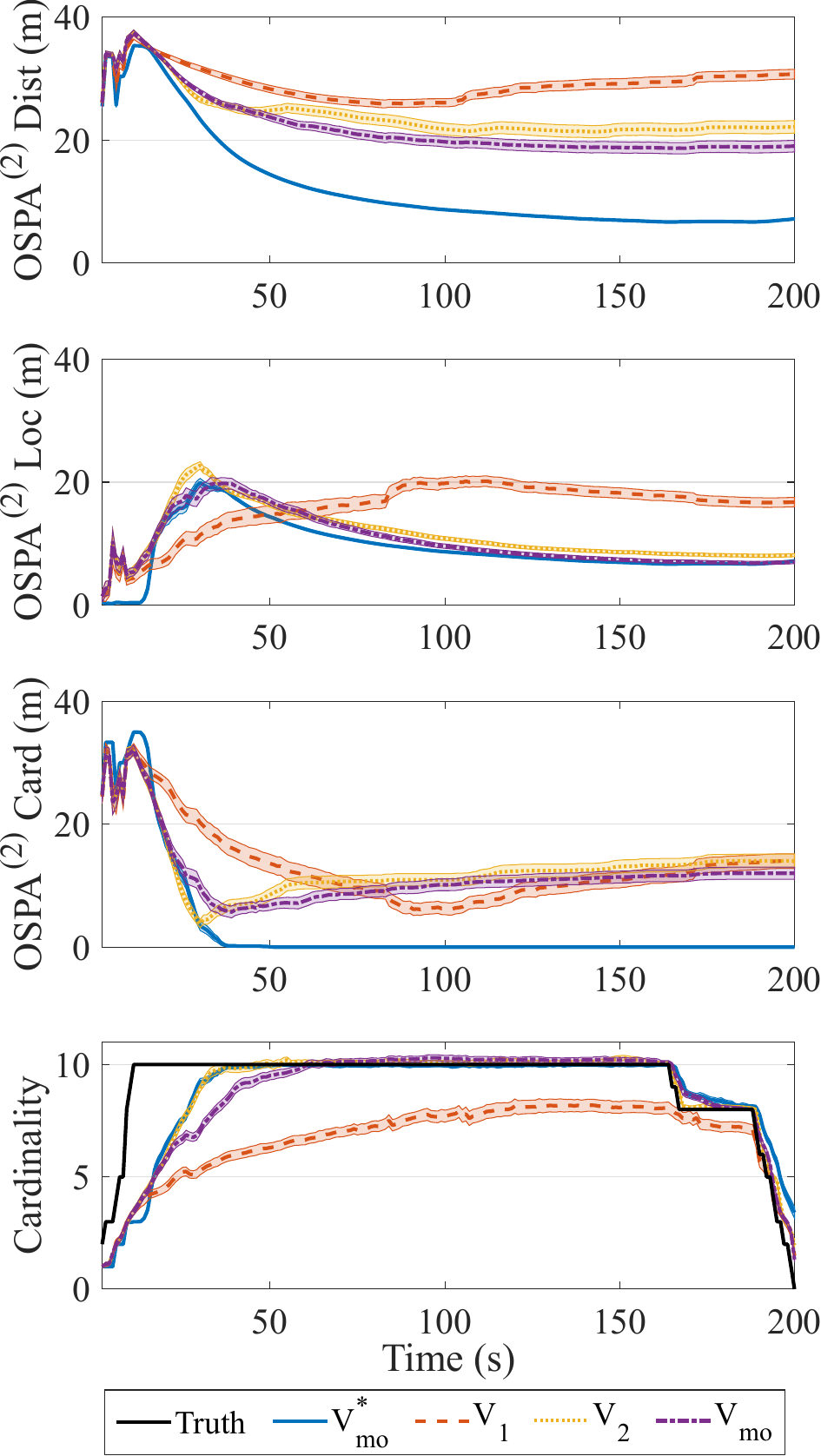}
\vspace{-0.2cm}
 \caption{Overall tracking performance of the 10 agents versus
time over $100$ MC runs (mean $\pm0.2$ standard deviation), for different planning
methods.}
\vspace{-0.4cm}
 \label{fig:scenario5_ospa2_and_card} 
\end{figure}

The mean and standard deviation (over $100$ MC runs) of the OSPA\textsuperscript{(2)}
and cardinality errors for the different planning methods in Fig.~\ref{fig:scenario5_ospa2_and_card},
shows that the proposed $V_{mo}$ consistently outperforms $V_{1}$
(tracking-only) or $V_{2}$ (discovering-only) in terms of the overall
tracking performance (\ie, OSPA\textsuperscript{(2)} Dist). Further,
for localisation (\ie, OSPA\textsuperscript{(2)} Loc), the performance
of $V_{mo}$ approaches that of the ideal case $V_{mo}^{*}$. We observe
that most of the planning methods (in the experiment) can correctly
estimate the number of taxis (\ie, Cardinality), except $V_{1}$,
which only focuses on tracking and does not explore the areas outside
the team's FoVs. In terms of OSPA\textsuperscript{(2)} Card, since
the cardinality is estimated correctly, most OSPA\textsuperscript{(2)}
Card errors come from the track label switching and track fragmentation,
which is expected given the limited FoVs, and the motion model mismatch
between the CT model and the taxi's occasional turns.

\subsection{Challenging Hypothetical Scenarios \label{subsec:Challenging-Hypothetical-Scenari}}

\begin{figure}[!tb]
\centering \includegraphics[width=0.45\textwidth]{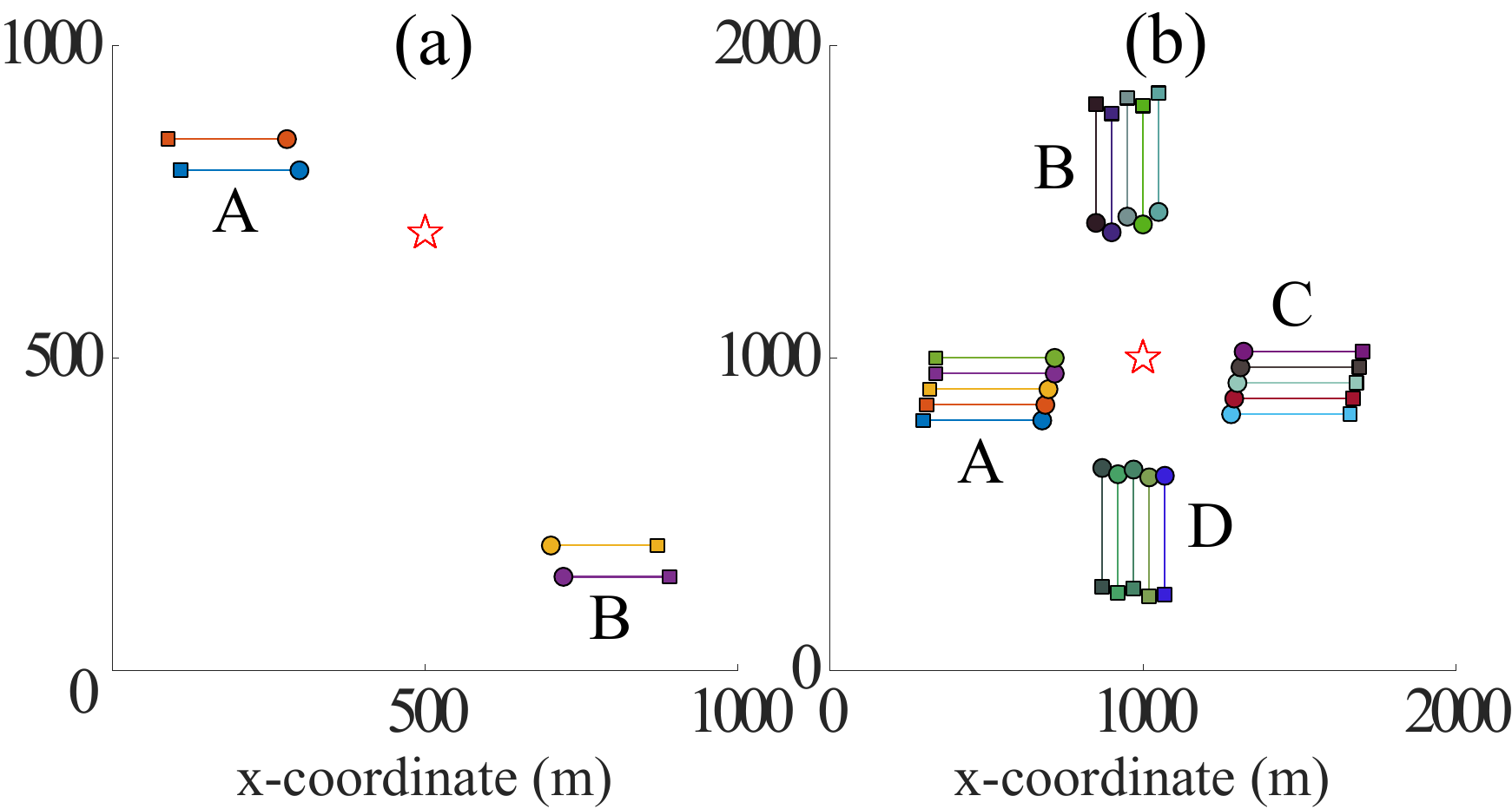}
\vspace{-0.3cm}
 \caption{Experimental settings for two simulated scenarios.
The red \textcolor{red}{$\star$} symbol denotes the agents' initial positions, while
the symbols $\bigcirc/\Box$ denote the Start/Stop positions of each
object. (a) Scenario 1: two groups, each containing 2 objects, in
a $1000$~m $\times1000$~m region, moving outwards in opposite
directions. Due to the sensor's limited detection range, locating
group $B$, positioned beyond the detection range, solely relies on
exploration. (b) Scenario 2: Four groups, each containing $5$ objects,
in a $2000$~m $\times2000$~m region, moving outwards rapidly in
opposite directions.}
\vspace{-0.3cm}
 \label{scenario_setting}
\end{figure}

To highlight the differences between the value functions, we test them on two challenging hypothetical
scenarios from \cite{hoa2020aaai}, illustrated in Fig.~\ref{scenario_setting}.

Similar to the previous experiment, the objects move
independently in a 2D plane, but according to a constant velocity
model $x_{k+1|k}=F^{CV}x_{k}+q_{k}^{CV}$, where $F^{CV}=[1,T_{0};0,T_{0}]\otimes I_{2}$,
$T_{0}=1$~s is the sampling interval, $\otimes$ is the Kronecker
product, $I_{2}$ is the $2\times2$ identity matrix, $q_{k}^{CV}\sim\mathcal{G}(\cdot;0,\Sigma^{CV})$
 is a 4D vector of zero-mean Gaussian process noise, with covariance
$\Sigma^{CV}=\sigma_{CV}^{2}[T_{0}^{3}/3,T_{0}^{2}/2;T_{0}^{2}/2,T_{0}]\otimes I_{2}$.
We also use the same \textit{adaptive birth}  model as the previous experiment, with a constant initial birth probability $r_{B,0}^{(i)}=0.01,~i=1:100$, and $\Delta_{x}=\Delta_{y}=100$~m for Scenario 1, and $\Delta_{x}=\Delta_{y}=200$~m for Scenario 2. A $100\times100$ occupancy grid (totalling $10,000$) cells) is used for both scenarios. Scenario 1 has a finer grid resolution with $10$~m$\times10$~m cells, while Scenario 2 has a coarser grid
with $20$~m$\times20$~m cells. 

\begin{table}[!tb]
	\centering
	\caption{Averaged tracking performance over 100 MC runs}
	\vspace{-0.2cm}
	\label{tab:scenario12}
	\begin{adjustbox}{max width=0.45\textwidth}
		\begin{tabular}{cl|rr||rr|}
			\cline{3-6}
			\multicolumn{1}{l}{}                       & \multicolumn{1}{l|}{} & \multicolumn{2}{c||}{\textbf{Scenario 1}} & \multicolumn{2}{c|}{\textbf{Scenario 2}} \\ \hline
			\multicolumn{1}{|c|}{\textbf{Agents}} &
			\textbf{\begin{tabular}[c]{@{}c@{}}Value \\ Func.\end{tabular}} &
			\multicolumn{1}{c|}{\textbf{\begin{tabular}[c]{@{}c@{}}OSPA$^{(2)}$ \\ Dist (m)\end{tabular}}} &
			\multicolumn{1}{c||}{\textbf{\begin{tabular}[c]{@{}c@{}}Entropy\\ (nats)\end{tabular}}} &
			\multicolumn{1}{c|}{\textbf{\begin{tabular}[c]{@{}c@{}}OSPA$^{(2)}$\\ Dist (m)\end{tabular}}} &
			\multicolumn{1}{c|}{\textbf{\begin{tabular}[c]{@{}c@{}}Entropy\\  (nats)\end{tabular}}} \\ \hline
			\multicolumn{1}{|c|}{\multirow{4}{*}{$|\mathcal{N}|=3$}} & $V_1$                    & \multicolumn{1}{r|}{29.0}     & 0.22     & \multicolumn{1}{r|}{34.1}     & 0.27     \\ \cline{2-2}
			\multicolumn{1}{|c|}{}                     & $V_2$                    & \multicolumn{1}{r|}{29.3}     & \textbf{0.06}     & \multicolumn{1}{r|}{35.5}     & \textbf{0.20}     \\ \cline{2-2}
			\multicolumn{1}{|c|}{}                     & $V_{mo}$                   & \multicolumn{1}{r|}{\textbf{26.5}}     & 0.12     & \multicolumn{1}{r|}{\textbf{30.1}}     & 0.23     \\ \cline{2-6} 
			\multicolumn{1}{|c|}{}                     & $V_{mo}^*$                  & \multicolumn{1}{r|}{8.2}      & 0.08     & \multicolumn{1}{r|}{17.2}     & 0.21     \\ \hline
			\multicolumn{1}{|c|}{\multirow{4}{*}{$|\mathcal{N}|=5$}} & $V_1$                    & \multicolumn{1}{r|}{28.3}     & 0.21     & \multicolumn{1}{r|}{32.6}     & 0.25     \\ \cline{2-2}
			\multicolumn{1}{|c|}{}                     & $V_2$                    & \multicolumn{1}{r|}{26.9}     & \textbf{0.03}     & \multicolumn{1}{r|}{36.6}     & \textbf{0.15}     \\ \cline{2-2}
			\multicolumn{1}{|c|}{}                     & $V_{mo}$                   & \multicolumn{1}{r|}{\textbf{26.3}}     & 0.06     & \multicolumn{1}{r|}{\textbf{26.9}}     & 0.19     \\ \cline{2-6} 
			\multicolumn{1}{|c|}{}                     & $V_{mo}^*$                  & \multicolumn{1}{r|}{6.0}      & 0.03     & \multicolumn{1}{r|}{11.4}     & 0.16     \\ \hline
		\end{tabular}
	\end{adjustbox}
\end{table}

\begin{figure}[!tb]
\vspace{-0.4cm}
\centering \includegraphics[width=0.44\textwidth]{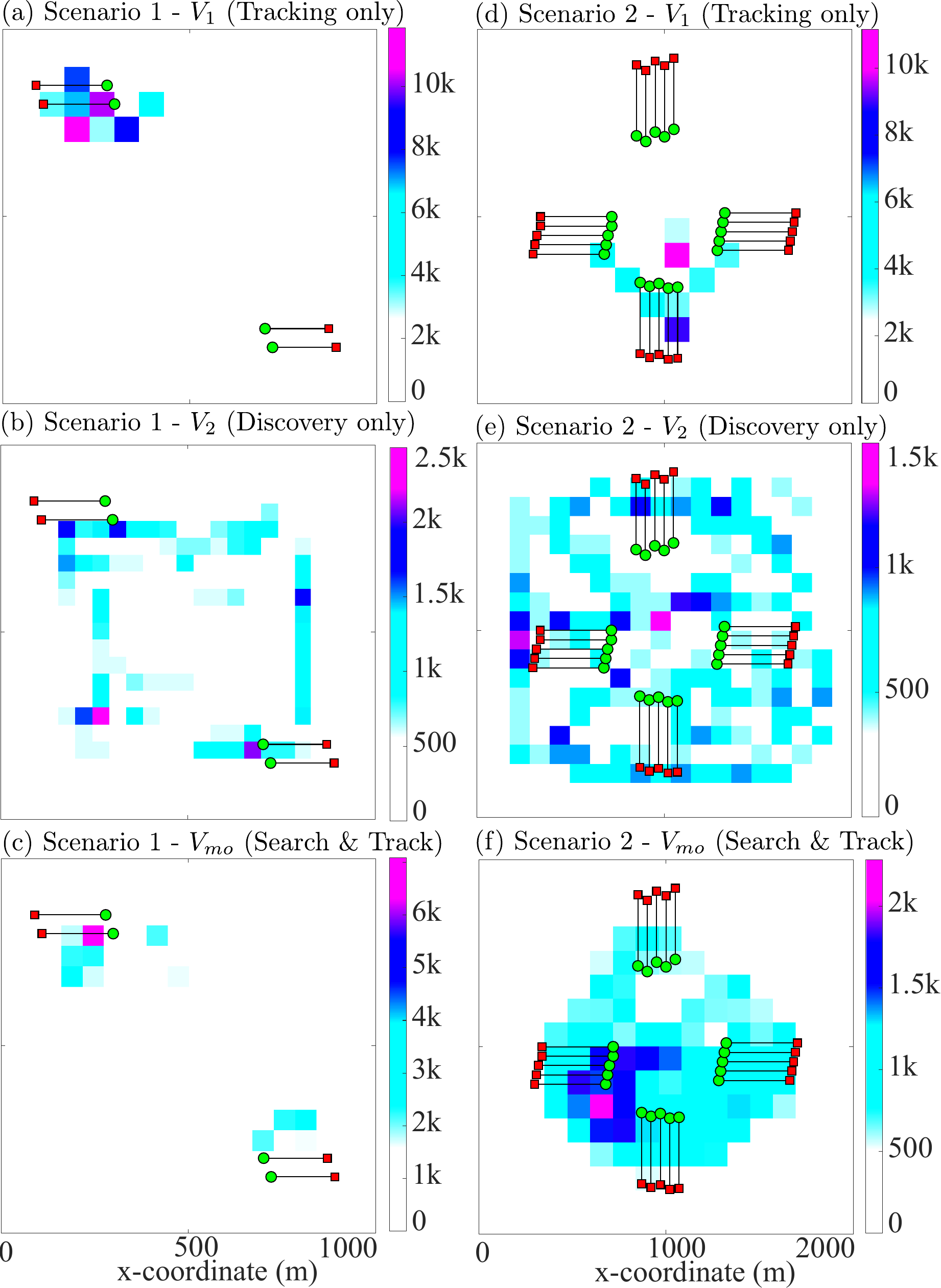}
\vspace{-0.3cm}
 \caption{Heatmap ($16\times16$ grids) showing the positions of $5$ UAVs and
objects trajectories over $100$ MC runs for Scenario 1 and Scenario
2. Circles ($\bigcirc$) and squares ($\Box$) indicate object start/stop
positions. The colour legend represents UAV position frequency.}
\vspace{-0.4cm}
 \label{fig:scenario34_heatmap}
\end{figure}

Similar to the previous experiment, to avoid collisions,
the agents operate at varying heights, with a minimum elevation of
$30$~m and a $5$~m vertical separation between drones. Each agent
has an on-board position sensor with a limited detection range of
$r_{D}=200$~m. The probability of detecting an object depends on its distance to the agent, and follows \eqref{eq:pd_by_distance}
with $P_{D}^{\max}=0.9$, $\hbar=0.008$~m$^{-1}$. When an object
$\boldsymbol{x}$ is detected, the sensor registers a noisy position
measurement $z=\big[\rho_{x},\rho_{y}]^{\dagger}+v$, where $v\sim\mathcal{G}(\cdot;0,R)$
is a 2D zero-mean Gaussian noise vector with covariance matrix $R=\mathrm{diag}([\sigma_{x}^{2},\sigma_{y}^{2}])$,
$\sigma_{x}=\sigma_{y}=5+0.01d(\boldsymbol{x},u^{(n)})$, to model
a distance-dependent uncertainty that increases as the object moves
away from the agent. In addition to the detected measurements, the
sensor also registers Poisson false alarms with rate $\lambda_{c}=15$.
The total simulation time is set to $200$~s. All reported results
are averaged over $100$ Monte Carlo runs.

Table~\ref{tab:scenario12} summarises performance
comparison for different value functions. The results indicate that
\textit{the proposed multi-objective value function},
$V_{mo}$, consistently outperforms the others in \textbf{OSPA}\textsuperscript{(2)}
tracking errors. This is because $V_{mo}$ strategically balances
between tracking already-detected objects and actively searching for
new ones. The average grid occupancy entropy (measured in nats) sheds
further insights into the coverage of the multi-agent team. A lower
entropy value indicates a lower level of uncertainty in discovering
new objects. As expected, $V_{2}$ exhibits the lowest entropy since
it focuses solely on discovering, while $V_{mo}$ balances between
tracking and discovery, achieving an entropy higher than that of $V_{2}$
but lower than $V_{1}$.

Fig. \ref{fig:scenario34_heatmap} shows the trajectory
heatmaps of five UAVs for both scenarios with different value functions.
These results further substantiate the efficacy of our proposed multi-objective
planning method. Specifically, for $V_{1}$ (tracking only; see Fig.
\ref{fig:scenario34_heatmap}(a) \& (d)), the agents prioritise solely tracking
detected objects and neglect any exploration tasks by only staying close to the detected objects in the northwestern region of Fig.
\ref{fig:scenario34_heatmap}(a), or in the southern region of Fig.
\ref{fig:scenario34_heatmap}(d). Conversely, $V_{2}$ (discovery only;
see Fig. \ref{fig:scenario34_heatmap}(b) \& (e)) drives the agents to focus
exclusively on exploration by circling the search area, disregarding
tracking duties. In contrast, $V_{mo}$ (joint search and track; see
Fig. \ref{fig:scenario34_heatmap}(c) \& (f)) encourages the agents to track
discovered objects and at the same time explore other parts of the
surveillance area to discover new objects. For instance, in Fig. \ref{fig:scenario34_heatmap}(c),
the agents move from the northwestern region to the southeastern region
to detect and track the two initially undetected objects. Similarly,
in Fig. \ref{fig:scenario34_heatmap}(f), the agents not only operate
in the southern region but also venture north to track the five initially
undetected objects.

\section{Conclusions \label{sec:conclusion}}

We have developed a multi-objective planning method for a limited-FoV
multi-agent team to discover and track multiple mobile objects with
unknown data association. The problem is formulated as a multi-agent
POMDP with the multi-object filtering density as the information state.
An entropy-based multi-objective value function is developed and shown
to be monotone and submodular, thereby enabling low-cost implementation
via greedy search with tight optimality bound. A series of experiments
on a real-world taxi dataset confirm our method's capability and efficiency.
It also validates the robustness of our proposed multi-objective formulation,
wherein its overall performance shows a decreasing trend over time
similar to that of the best possible performance. So far, we have
considered a centralised architecture for multi-agent planning where
scalability can be a limitation. A scalable approach should be a distributed
POMDP for MOT, where each agent runs its local filter to track multi-objects 
and coordinates with other agents to achieve a global objective. However,
planning for multi-agents to reach a global goal under a distributed
POMDP framework is an NEXP-complete problem~\cite{bernstein2002complexity}.

\bibliographystyle{IEEEtran}
\bibliography{references}

\appendix
\section{}

\subsection{MS-GLMB Recursion}

\label{sec:ms_glmb_appendix} 
Within the conventional multi-object framework, the
MS-GLMB recursion, denoted by $\Omega$, propagates the GLMB parameter
set:
\[
\boldsymbol{\pi}\triangleq\left\{ \left(w^{\left(\xi,I\right)},p^{\left(\xi\right)}\right):\left(\xi,I\right)\in\Xi\times\mathcal{F}(\mathbb{L})\right\}, 
\]
to the parameter set ~\cite{vo2019multisensor}
\[
\boldsymbol{\pi}_{+}=\left\{ \left(w_{+}^{\left(\xi_{+},I_{+}\right)},p_{+}^{\left(\xi_{+}\right)}\right)\!:\left(\xi_{+},I_{+}\right)\in\Xi_{+}\times\mathcal{F}(\mathbb{L_{+}})\right\} ,
\]
where:
\begin{equation}
I_{+}=I\uplus\mathbb{B}_{+},\;\xi_{+}=(\xi,\gamma_{+}),\label{e:GLMB_JPU_First-1-1-1-1}
\end{equation}
\begin{equation}
w_{+}^{\left(I_{+},\xi_{+}\right)}=1_{\mathcal{F}(\mathbb{B}_{+}\uplus I)}(\mathcal{L}(\gamma_{+}))w^{(I,\xi)}\left[\omega^{(\xi,\gamma_{+})}\right]^{\mathbb{B}_{+}\uplus I\!},\label{eq:Upda_weight}
\end{equation}
\begin{multline}
\!\!\!\!\!\!\!p_{+}^{\left(\xi_{+}\right)}(x_{+},\ell)\propto\\
\begin{cases}
\langle p^{(\xi)}(\cdot,\ell),\Upsilon_{S}^{(\gamma_{+})}(x_{+}|\cdot,\ell)\rangle, & \!\!\ell\in\mathcal{L}(\gamma_{+})\setminus\mathbb{B_{+}}\\
\Upsilon_{B}^{(\gamma_{+})}(x_{+},\ell), & \!\!\ell\in\mathcal{L}(\gamma_{+})\cap\mathbb{B_{+}}
\end{cases},\label{eq:_plus_p-1-1}
\end{multline}
\begin{equation}
\!\!\!\!\!\!\!\!\!\!\!\!\!\!\!\omega^{(\xi,\gamma_{+})}(\ell)=\begin{cases}
1-\bar{r}_{S}^{\left(\xi\right)}\left(\ell\right), & \!\!\ell\in\overline{\mathcal{L}(\gamma_{+})}\setminus\mathbb{B_{+}}\\
\overline{\Upsilon}_{S}^{(\xi,\gamma_{+})}(\ell), & \!\!\ell\in\mathcal{L}(\gamma_{+})\setminus\mathbb{B_{+}}\\
1-r_{B,+}(\ell), & \!\!\ell\in\overline{\mathcal{L}(\gamma_{+})}\cap\mathbb{B_{+}}\\
\overline{\Upsilon}_{B}^{(\gamma_{+})}(\ell), & \!\!\ell\in\mathcal{L}(\gamma_{+})\cap\mathbb{B}_{+}
\end{cases},
\end{equation}
\begin{align}
\bar{r}_{S}^{\left(\xi\right)}\left(\ell\right)= & \langle p^{(\xi)}(\cdot,\ell),r_{S}(\cdot,\ell)\rangle,\\
\!\!\!\Upsilon_{B}^{(\gamma_{+})}(x_{+},\ell)= & p_{B,+}(x_{+},\ell)r_{B,+}(\ell)\psi^{(\gamma_{+})}~\!\!\!\left(x_{+},\ell\right),\\
\!\!\!\!\Upsilon_{S}^{(\gamma_{+})}(x_{+}|y,\ell)= & f_{+}(x_{+}|y,\ell)r_{S}(y,\ell)\psi^{(\gamma_{+})}~\!\!\!\left(x_{+},\ell\right),\\
\overline{\Upsilon}_{B}^{(\gamma_{+})}(\ell)= & \int\Upsilon_{B}^{(\gamma_{+})}(x,\ell)dx,\\
\overline{\Upsilon}_{S}^{(\xi,\gamma_{+})}(\ell)= & \int\langle p^{(\xi)}(\cdot,\ell),\Upsilon_{S}^{(\gamma_{+})}(x|\cdot,\ell)\rangle dx,
\end{align}
\begin{align}
\psi^{(\gamma)}\left(\boldsymbol{x}\right) & \triangleq\prod\limits _{n\in\mathcal{N}}\psi^{(\gamma^{\left(n\right)},n)}\left(\boldsymbol{x}\right),\label{e:MS_Lkhd_Abbrev_6-1-1}
\end{align}
\begin{align*}
\!\psi^{(\gamma^{\left(n\right)},n)}\!\negthinspace\left(\boldsymbol{x}\right)\negthinspace=\negthinspace & \begin{cases}
\!1-P_{D}^{\left(n\right)}\!\left(\boldsymbol{x}\right)\!,\!\!\!\! & \negthinspace\!\!\!\!\gamma^{\left(n\right)}(\mathcal{L}(\boldsymbol{x}))\negthinspace=\negthinspace0\\
\!\frac{P_{D}^{\left(n\right)\!}\left(\boldsymbol{x}\right)g^{\left(n\right)\!}({z}_{j}^{(n)}|\boldsymbol{x})}{\kappa^{\left(a\right)}(z_{j}^{(n)})}\!, & \negthinspace\!\!\!\!\gamma^{\left(n\right)}(\mathcal{L}(\boldsymbol{x}))\negthinspace=\negthinspace j\negthinspace\!>\negthinspace0
\end{cases}\negthinspace.\!\!
\end{align*}

\subsection{Analytical Form of Tracking Value Function \label{subsec:Proof-of-Tracking}}

To prove Proposition \textbf{\ref{prop:diff_entropy_lmb}}, we first
require the three following Lemmas.

\begin{lemma} \label{lemma:1} For $f:\mathcal{F}(\mathbb{L})\rightarrow\mathbb{R}$,
and $g:\mathbb{L}\rightarrow\mathbb{R}$,: 
\begin{equation}
\sum_{L\subseteq\mathbb{L}}f(L)\sum_{\ell\in L}g(\ell)=\sum_{\ell\in\mathbb{L}}g(\ell)\sum_{L\subseteq\mathbb{L}\setminus\{\ell\}}f(L\cup\{\ell\})\label{eq:lemma_1}
\end{equation}
\end{lemma}

\textbf{Proof:} 
Notably, a specific case ($f(L)=[1-r_{1}^{(\ell)}]^{\mathbb{L}\setminus L}[r_{1}^{(\ell)}]^{L}$
and $g(\ell)=\ln{({r_{1}^{(\ell)}}/{r_{2}^{(\ell)}})}$) of this result
was first used in Proposition 1 of \cite{shen2022consensus} without
proof. This lemma presents a more general result and is proven via induction. 

It is trivial to confirm that \eqref{eq:lemma_1} holds for $\mathbb{L}=\emptyset$
and $\mathbb{L}=\{\ell\}$. Suppose \eqref{eq:lemma_1} holds for
$\mathbb{L}=\{\ell_{1},\dots,\ell_{n}\}$. It remains to show that
\eqref{eq:lemma_1} also holds for $\mathbb{L}\cup\{\hat{\ell}\}$,
i.e., \resizebox{0.999\linewidth}{!}{ %
\noindent\begin{minipage}[c]{1\linewidth}%
\begin{alignat}{1}
\!\!\!\!\!\!\sum_{L\subseteq\mathbb{L}\cup\hat{\ell}}f(L)\sum_{\ell\in L}g(\ell) & =\sum_{\ell\in\mathbb{L}\cup\hat{\ell}}g(\ell)\sum_{L\subseteq\mathbb{L}\cup\hat{\ell}\setminus\{\ell\}}f(L\cup\{\ell\}).\label{eq:lemma_1_lnp1}
\end{alignat}
\end{minipage}}

Note that for any function $f:\mathcal{F}(\mathbb{L})\rightarrow\mathbb{R}$,
\resizebox{0.95\linewidth}{!}{ %
\noindent\begin{minipage}[c]{1\linewidth}%
\begin{alignat}{1}
\sum_{L\subseteq\mathbb{L}\cup\hat{\ell}}f(L)=\sum_{L\subseteq\mathbb{L}}f(L)+\sum_{L\subseteq\mathbb{L}}f(L\cup\{\hat{\ell}\})\label{eq:27}
\end{alignat}
\end{minipage}} since $\mathcal{F}(\mathbb{L}\cup\hat{\ell})=\mathcal{F}(\mathbb{L})\cup\{(L\cup\{\widehat{\ell}\}):L\in\mathcal{F}(\mathbb{L})\}$
where $\mathcal{F}(\cdot)$ denotes the class of finite subsets. Hence,
\resizebox{0.95\linewidth}{!}{ %
\noindent\begin{minipage}[c]{1\linewidth}%
\begin{alignat}{1}
\!\!\!\!\!\!\!\!\!\! & \sum_{L\subseteq\mathbb{L}\cup\hat{\ell}}f(L)\sum_{\ell\in L}g(\ell)\nonumber \\
= & \sum_{L\subseteq\mathbb{L}}f(L)\sum_{\ell\in L}g(\ell)+\sum_{L\subseteq\mathbb{L}}f(L\cup\{\hat{\ell}\})\sum_{\ell\in L\cup\{\hat{\ell}\}}g(\ell)\label{eq:28}
\end{alignat}
\end{minipage}}

Substituting \eqref{eq:lemma_1} into \eqref{eq:28}, and noting that
$\sum_{\ell\in L\cup\{\hat{\ell}\}}g(\ell)=\sum_{\ell\in L}g(\ell)+g(\hat{\ell})$,
we have: \resizebox{0.95\linewidth}{!}{ %
\noindent\begin{minipage}[c]{1\linewidth}%
\begin{alignat*}{1}
 & \sum_{L\subseteq\mathbb{L}\cup\hat{\ell}}f(L)\sum_{\ell\in L}g(\ell)\\
= & \sum_{\ell\in\mathbb{L}}g(\ell)\sum_{L\subseteq\mathbb{L}\setminus\{\ell\}}f(L\cup\{\ell\})+\sum_{L\subseteq\mathbb{L}}f(L\cup\{\hat{\ell}\})\sum_{\ell\in L}g(\ell)\\
 & +\sum_{L\subseteq\mathbb{L}}f(L\cup\{\hat{\ell}\})g(\hat{\ell})\\
= & \sum_{\ell\in\mathbb{L}}g(\ell)\sum_{L\subseteq\mathbb{L}\setminus\{\ell\}}f(L\cup\{\ell\})\\
 & +\sum_{\ell\in\mathbb{L}}g(\ell)\sum_{L\subseteq\mathbb{L}\setminus\{\ell\}}f(L\cup\{\hat{\ell}\}\cup\{\ell\})+g(\hat{\ell})\sum_{L\subseteq\mathbb{L}}f(L\cup\{\hat{\ell}\})\\
= & \sum_{\ell\in\mathbb{L}}g(\ell)\sum_{L\subseteq\mathbb{L}\cup\hat{\ell}\setminus\{\ell\}}f(L\cup\{\ell\})+g(\hat{\ell})\sum_{L\subseteq\mathbb{L}}f(L\cup\{\hat{\ell}\})\\
= & \sum_{\ell\in\mathbb{L}\cup\hat{\ell}}g(\ell)\sum_{L\subseteq\mathbb{L}\cup\hat{\ell}\setminus\{\ell\}}f(L\cup\{\ell\})\quad\Box.
\end{alignat*}
\end{minipage}}

\vspace{4mm}
 \begin{lemma} \label{lemma:2} For $f:\mathcal{F}(\mathbb{L})\rightarrow\mathbb{R}$,
and $g:\mathbb{L}\rightarrow\mathbb{R}$, we have: 
\begin{equation}
\sum_{L\subseteq\mathbb{L}}f(L)\sum_{\ell\in\mathbb{L}\setminus L}g(\ell)=\sum_{\ell\in\mathbb{L}}g(\ell)\sum_{L\subseteq\mathbb{L}\setminus\{\ell\}}f(L).\label{eq:lemma_2}
\end{equation}
\end{lemma}

\textbf{Proof: } We prove this Lemma via induction. It is clear that
\eqref{eq:lemma_2} holds for $\mathbb{L}=\emptyset$ and $\mathbb{L}=\{\ell\}$.
Now, assuming \eqref{eq:lemma_2} for $\mathbb{L}=\{\ell_{1},\dots,\ell_{n}\}$,
we demonstrate its validity for $\mathbb{L}\cup{\hat{\ell}}$, i.e.,:
\begin{align}
\sum_{L\subseteq\mathbb{L}\cup\hat{\ell}}f(L)\sum_{\ell\in\mathbb{L}\cup\hat{\ell}\setminus L}g(\ell) & =\sum_{\ell\in\mathbb{L}\cup\hat{\ell}}g(\ell)\sum_{L\subseteq\mathbb{L}\cup\hat{\ell}\setminus\{\ell\}}f(L).\label{eq:lemma_2_lpn1}
\end{align}

\begin{flushleft}
Using \eqref{eq:27} and noting that $\sum_{\ell\in\mathbb{L}\cup\{\hat{\ell}\}\setminus L}g(\ell)$
$=\sum_{\ell\in\mathbb{L}\setminus L}g(\ell)$ $+$ $g(\hat{\ell})$,
we have: 
\par\end{flushleft}

\resizebox{0.99\linewidth}{!}{ %
\noindent\begin{minipage}[c]{1\linewidth}%
\begin{alignat}{1}
 & \sum_{L\subseteq\mathbb{L}\cup\hat{\ell}}f(L)\sum_{\ell\in\mathbb{L}\cup\hat{\ell}\setminus L}g(\ell)\nonumber \\
= & \sum_{L\subseteq\mathbb{L}}f(L)\sum_{\ell\in\mathbb{L}\cup\hat{\ell}\setminus L}g(\ell)+\sum_{L\subseteq\mathbb{L}}f(L\cup\hat{\ell})\sum_{\ell\in\mathbb{L}\setminus L}g(\ell)\nonumber \\
= & \sum_{L\subseteq\mathbb{L}}f(L)\sum_{\ell\in\mathbb{L}\setminus L}g(\ell)+\sum_{L\subseteq\mathbb{L}}f(L)g(\hat{\ell})+\sum_{L\subseteq\mathbb{L}}f(L\cup\hat{\ell})\sum_{\ell\in\mathbb{L}\setminus L}g(\ell).\label{eq:31}
\end{alignat}
\end{minipage}} 
\begin{flushleft}
Substituting \eqref{eq:lemma_2} into \eqref{eq:31}, we have: 
\par\end{flushleft}

\resizebox{0.99\linewidth}{!}{ %
\noindent\begin{minipage}[c]{1\linewidth}%
\begin{alignat*}{1}
 & \sum_{L\subseteq\mathbb{L}\cup\hat{\ell}}f(L)\sum_{\ell\in\mathbb{L}\cup\hat{\ell}\setminus L}g(\ell)\\
= & \sum_{\ell\in\mathbb{L}}g(\ell)\sum_{L\subseteq\mathbb{L}\setminus\{\ell\}}f(L)+g(\hat{\ell})\sum_{L\subseteq\mathbb{L}}f(L)+\sum_{\ell\in\mathbb{L}}g(\ell)\sum_{L\subseteq\mathbb{L}\setminus\{\ell\}}f(L\cup\hat{\ell})\\
= & \sum_{\ell\in\mathbb{L}}g(\ell)\sum_{L\subseteq\mathbb{L}\cup\hat{\ell}\setminus\{\ell\}}f(L)+g(\hat{\ell})\sum_{L\subseteq\mathbb{L}\cup\hat{\ell}\setminus\{\ell\}}f(L)\\
= & \sum_{\ell\in\mathbb{L}\cup\hat{\ell}}g(\ell)\sum_{L\subseteq\mathbb{L}\cup\hat{\ell}\setminus\{\ell\}}f(L)\quad\Box.
\end{alignat*}
\end{minipage}}

\vspace{4mm}
\begin{lemma} \label{lemma:3} For $f:\mathcal{F}(\mathbb{L})\rightarrow\mathbb{R}$,
$p:\mathbb{X}\times\mathbb{L}\rightarrow\mathbb{R}$, $q:\mathbb{X}\times\mathbb{L}\rightarrow\mathbb{R}$
with $q$ is a unitless function, we have: 
\[
\int\Delta(\boldsymbol{X})f(\mathcal{L}(\boldsymbol{X}))p^{\boldsymbol{X}}\ln q^{\boldsymbol{X}}\delta\boldsymbol{X}=\sum_{L\subseteq\mathbb{L}}f(L)\left\langle p\right\rangle ^{L}\sum_{\ell\in L}\frac{\left\langle p\ln q\right\rangle (\ell)}{\left\langle p\right\rangle (\ell)}.
\]
\end{lemma} 
\begin{flushleft}
\textbf{Proof}: We first note that 
\par\end{flushleft}

\resizebox{0.99\linewidth}{!}{ %
\noindent\begin{minipage}[c]{1\linewidth}%
\begin{alignat}{1}
I & \triangleq\int\prod_{j=1}^{i}p(x_{j},\ell_{j})\ln\left(\prod_{n=1}^{i}q(x_{n},\ell_{n})\right)dx_{1:i}\nonumber \\
 & =\int\prod_{j=1}^{i}p(x_{j},\ell_{j})\sum_{n=1}^{i}\ln q(x_{n},\ell_{n})dx_{1:i}\nonumber \\
 & =\sum_{n=1}^{i}\int p(x,\ell_{n})\ln q(x,\ell_{n})dx\!\!\prod_{j\in\{1,\dots,i\}\setminus\{n\}}\!\!\int p(x,\ell_{j})dx\nonumber \\
 & =\sum_{\ell\in\{\ell_{1:i}\}}\left\langle p\ln q\right\rangle (\ell)\left\langle p\right\rangle ^{\{\ell_{1:i}\}-\{\ell\}}.~\label{eq:36}
\end{alignat}
\end{minipage}} Now 
\begin{align}
 & \int\Delta(\boldsymbol{X})f(\mathcal{L}(\boldsymbol{X}))p^{\boldsymbol{X}}\ln q^{\boldsymbol{X}}\delta\boldsymbol{X}\nonumber \\
 & =f(\emptyset)p^{\emptyset}\ln q^{\emptyset}+\sum_{i=1}^{\infty}\frac{1}{i!}\sum_{\ell_{1:i}}\delta_{i}[|\{\ell_{1:i}\}|]f(\{\ell_{1},\dots,\ell_{i}\})\times I,\nonumber \\
 & =\sum_{i=1}^{\infty}\frac{1}{i!}\sum_{\ell_{1:i}}\delta_{i}[|\{\ell_{1:i}\}|]f(\{\ell_{1},\dots,\ell_{i}\})\times I,\label{eq:35}\\
 & =\sum_{i=1}^{\infty}\frac{1}{i!}i!\sum_{L:|L|=i}f(L)\left[\sum_{\ell\in L}\left\langle p\right\rangle ^{L-\{\ell\}}\left\langle p\ln q\right\rangle (\ell)\right]\label{eq:35a}\\
 & =\sum_{L\subseteq\mathbb{L}}f(L)\left\langle p\right\rangle ^{L}\sum_{\ell\in L}\frac{\left\langle p\ln q\right\rangle (\ell)}{\left\langle p\right\rangle (\ell)}
\end{align}
where \eqref{eq:35a} follows by substituting \eqref{eq:36} into
\eqref{eq:35}, and the last step follows from Lemma 12 in \cite{vo2013glmb},
and noting that 
\begin{align}
f(\emptyset)\left\langle p\right\rangle ^{\emptyset}\sum_{\ell\in\emptyset}\frac{\left\langle p\ln q\right\rangle (\ell)}{\left\langle p\right\rangle (\ell)}=0~~~\Box.\hspace{2.5cm}
\end{align}

\begin{flushleft}
\textbf{Proof of Proposition \ref{prop:diff_entropy_lmb}: } For the
LMB density, we have $\boldsymbol{\pi}(\boldsymbol{X})=\Delta(\boldsymbol{X})\tilde{r}^{\mathbb{L}\setminus\mathcal{L}(\boldsymbol{X})}r^{\mathcal{L}(\boldsymbol{X})}p^{\boldsymbol{X}}$;
hence: 
\par\end{flushleft}

\resizebox{0.999\linewidth}{!}{ %
\noindent\begin{minipage}[c]{1\linewidth}%

\begin{alignat}{1}
- & h(\boldsymbol{X})=\int\boldsymbol{\pi}(\boldsymbol{X})\ln\big(K^{|\boldsymbol{X}|}\boldsymbol{\pi}(\boldsymbol{X})\big)\delta\boldsymbol{X}\nonumber \\
= & \int\Delta(\boldsymbol{X})\tilde{r}^{\mathbb{L}\setminus\mathcal{L}(\boldsymbol{X})}r^{\mathcal{L}(\boldsymbol{X})}p^{\boldsymbol{X}}\ln\big[\Delta(\boldsymbol{X})\tilde{r}^{\mathbb{L}\setminus\mathcal{L}(\boldsymbol{X})}r^{\mathcal{L}(\boldsymbol{X})}p^{\boldsymbol{X}}K^{|\boldsymbol{X}|}\big]\delta\boldsymbol{X}\nonumber \\
= & \int\Delta(\boldsymbol{X})\tilde{r}^{\mathbb{L}\setminus\mathcal{L}(\boldsymbol{X})}r^{\mathcal{L}(\boldsymbol{X})}p^{\boldsymbol{X}}\ln\big[\tilde{r}^{\mathbb{L}\setminus\mathcal{L}(\boldsymbol{X})}r^{\mathcal{L}(\boldsymbol{X})}\big]\delta\boldsymbol{X}\nonumber \\
 & +\int\Delta(\boldsymbol{X})\tilde{r}^{\mathbb{L}\setminus\mathcal{L}(\boldsymbol{X})}r^{\mathcal{L}(\boldsymbol{X})}p^{\boldsymbol{X}}\ln\big[p^{\boldsymbol{X}}K^{|\boldsymbol{X}|}\big]\delta\boldsymbol{X}.\label{eq:diff_entropy_lmb_2_terms}
\end{alignat}
\end{minipage}}
\begin{flushleft}
The first term on the right-hand side (RHS) of \eqref{eq:diff_entropy_lmb_2_terms}
is: 

\par\end{flushleft}

\resizebox{0.95\linewidth}{!}{ %
\noindent\begin{minipage}[c]{1\linewidth}%

\begin{alignat}{1}
 & \int\Delta(\boldsymbol{X})\tilde{r}^{\mathbb{L}\setminus\mathcal{L}(\boldsymbol{X})}r^{\mathcal{L}(\boldsymbol{X})}p^{\boldsymbol{X}}\ln\big[\tilde{r}^{\mathbb{L}\setminus\mathcal{L}(\boldsymbol{X})}r^{\mathcal{L}(\boldsymbol{X})}\big]\delta\boldsymbol{X}\\
= & \sum_{L\subseteq\mathbb{L}}\tilde{r}^{\mathbb{L}\setminus L}r^{L}\ln\big[\tilde{r}^{\mathbb{L}\setminus L}r^{L}\big]\bigg[\int p(x,\cdot)dx\bigg]^{L}\label{eq:first_term_eq1}\\
= & \sum_{L\subseteq\mathbb{L}}\tilde{r}^{\mathbb{L}\setminus L}r^{L}\sum_{\ell\in\mathbb{L}\setminus L}\ln\tilde{r}^{(\ell)}+\sum_{L\subseteq\mathbb{L}}\tilde{r}^{\mathbb{L}\setminus L}r^{L}\sum_{\ell\in L}\ln r^{(\ell)}\label{eq:first_term_eq2}\\
= & \sum_{\ell\in\mathbb{L}}\ln\tilde{r}^{(\ell)}\sum_{L\subseteq\mathbb{L}\setminus\{\ell\}}\tilde{r}^{\mathbb{L}\setminus L}r^{L}\nonumber \\
 & \quad+\sum_{\ell\in\mathbb{L}}\ln r^{(\ell)}\sum_{L\subseteq\mathbb{L}\setminus\{\ell\}}\tilde{r}^{\mathbb{L}\setminus(L\cup\{\ell\})}r^{L\cup\{\ell\}}\label{eq:first_term_eq3}\\
= & \sum_{\ell\in\mathbb{L}}\bigg[\tilde{r}^{(\ell)}\ln\tilde{r}^{(\ell)}\sum_{L\subseteq\mathbb{L}\setminus\{\ell\}}\tilde{r}^{(\mathbb{L}\setminus\{\ell\})\setminus L}r^{L}\nonumber \\
 & \quad\quad\quad+r^{(\ell)}\ln r^{(\ell)}\sum_{L\subseteq\mathbb{L}\setminus\{\ell\}}\tilde{r}^{(\mathbb{L}\setminus\{\ell\})\setminus L}r^{L}\bigg]\label{eq:first_term_eq4}\\
= & \sum_{\ell\in\mathbb{L}}\left[r^{(\ell)}\ln r^{(\ell)}+\tilde{r}^{(\ell)}\ln\tilde{r}^{(\ell)}\right].\label{eq:first_term_eq5}
\end{alignat}
\end{minipage}} \vspace{0.1cm}

Here, \eqref{eq:first_term_eq1} follows from Lemma 3 in \cite{vo2013glmb};
\eqref{eq:first_term_eq2} follows from $\int p(x,\cdot)dx=1$; each
term in \eqref{eq:first_term_eq3} follows from Lemma~\ref{lemma:2}
and Lemma~\ref{lemma:1}, respectively; while \eqref{eq:first_term_eq5}
follows from the Binomial Theorem, i.e., $\sum_{L\subseteq\mathbb{L}}f^{\mathbb{L}\setminus L}g^{L}=(f+g)^{\mathbb{L}}$.
Hence, $\sum_{L\subseteq\mathbb{L}\setminus\{\ell\}}\tilde{r}^{(\mathbb{L}\setminus\{\ell\})\setminus L}r^{L}=(\tilde{r}+r)^{\mathbb{L}\setminus\{\ell\}}=1$.

\begin{flushleft}
The second term of the RHS of \eqref{eq:diff_entropy_lmb_2_terms}
is: 
\par\end{flushleft}

\resizebox{0.95\linewidth}{!}{ %
\noindent\begin{minipage}[c]{1\linewidth}%
\begin{alignat}{1}
 & \int\Delta(\boldsymbol{X})\tilde{r}^{\mathbb{L}\setminus\mathcal{L}(\boldsymbol{X})}r^{\mathcal{L}(\boldsymbol{X})}p^{\boldsymbol{X}}\ln\big[p^{\boldsymbol{X}}K^{|\boldsymbol{X}|}\big]\delta\boldsymbol{X}\\
= & \sum_{L\subseteq\mathbb{L}}\tilde{r}^{\mathbb{L}\setminus L}r^{L}\sum_{\ell\in L}\left\langle p^{(\ell)}\ln(Kp^{(\ell)})\right\rangle \label{eq:second_term_eq1}\\
= & \sum_{\ell\in\mathbb{L}}\left\langle p^{(\ell)}\ln(Kp^{(\ell)})\right\rangle \sum_{L\subseteq\mathbb{L}\setminus\{\ell\}}\tilde{r}^{\mathbb{L}\setminus(L\cup\{\ell\})}r^{L\cup\{\ell\}}\label{eq:second_term_eq2}\\
= & \sum_{\ell\in\mathbb{L}}r^{(\ell)}\left\langle p^{(\ell)}\ln(Kp^{(\ell)})\right\rangle \sum_{L\subseteq\mathbb{L}\setminus\{\ell\}}\tilde{r}^{(\mathbb{L}\setminus\{\ell\})\setminus L}r^{L}\label{eq:second_term_eq3}\\
= & \sum_{\ell\in\mathbb{L}}r^{(\ell)}\left\langle p^{(\ell)}\ln(Kp^{(\ell)})\right\rangle .\label{eq:second_term_eq4}
\end{alignat}
\end{minipage}} \vspace{0.2cm}

\noindent Here, \eqref{eq:second_term_eq1} follows from Lemma~\ref{lemma:3};
\eqref{eq:second_term_eq2} follows from Lemma~\ref{lemma:1}; while
\eqref{eq:second_term_eq4} follows from the Binomial Theorem, i.e.,
$\sum_{L\subseteq\mathbb{L}\setminus\{\ell\}}\tilde{r}^{(\mathbb{L}\setminus\{\ell\})\setminus L}r^{L}=(\tilde{r}+r)^{\mathbb{L}\setminus\{\ell\}}=1$.

\vspace{0.2cm}
 Substituting \eqref{eq:first_term_eq5} and \eqref{eq:second_term_eq4}
into \eqref{eq:diff_entropy_lmb_2_terms}, we have: 
\begin{gather*}
h(\boldsymbol{X})=-\sum_{\ell\in\mathbb{L}}\left[r^{(\ell)}\ln r^{(\ell)}+\tilde{r}^{(\ell)}\ln\tilde{r}^{(\ell)}+r^{(\ell)}\left\langle p^{(\ell)}\ln(Kp^{(\ell)})\right\rangle \right]\Box.
\end{gather*}

\vspace{-0.4cm}

\subsection{Analytical Form of Discovery Value Function \label{subsec:Proof-of-Discovery}}

\textbf{Proof of Proposition \ref{prop:diff_entropy_occ}: }The predicted
occupancy probability in \eqref{eq:occ-pred} is straight forward.
To establish \eqref{eq:occ-up}, note that 
\begin{eqnarray*}
\textrm{Pr}(O_{j+1}^{(i)}=1,Y_{a_{j},j+1}^{(i)}=1)\! & =\! & \omega_{j+1}^{(i)}(a_{j-1})Q_{D,+}^{(i)}(a_{j}),\\
\textrm{Pr}(O_{j+1}^{(i)}=0,Y_{a_{j},j+1}^{(i)}=1)\! & =\! & 1-\omega_{j+1}^{(i)}(a_{j-1}).
\end{eqnarray*}

\begin{flushleft}
Therefore, $\textrm{Pr}(Y_{a_{j},j+1}^{(i)}=1)=1-\omega{}_{j+1}^{(i)}(a_{j-1})+$
$\omega_{j+1}^{(i)}(a_{j-1})Q_{D,+}^{(i)}(a_{j})$, and using Bayes'
rule yields \eqref{eq:occ-up}. 
\par\end{flushleft}

Note that if $\varkappa^{(i)}$ generates measurements (i.e., $Y_{a,j}^{(i)}=0$),
then it is occupied (i.e., $O_{j}^{(i)}=1$), which means 
\begin{eqnarray*}
\textrm{Pr}(O_{j}^{(i)}=0|Y_{a,j}^{(i)}=0)\! & =\! & 0.\\
\textrm{Pr}(O_{j}^{(i)}=1|Y_{a,j}^{(i)}=0)\! & =\! & 1.
\end{eqnarray*}
Hence, $h(O_{j}^{(i)}|Y_{a,j}^{(i)}=0)=0\ln0+1\ln1=0$, and 
\begin{align}
h(O_{j}^{(i)}|Y_{a,j}^{(i)}) & =\textrm{Pr}(Y_{a,j}^{(i)}=1)h(O_{j}^{(i)}|Y_{a,j}^{(i)}=1).\label{eq:diff_occ_cond_entropy}
\end{align}
Substituting for $\textrm{Pr}(O_{j}^{(i)}=1|Y_{a,j}^{(i)}=1)$ and
$\textrm{Pr}(Y_{a,j}^{(i)}=1)$ yields \eqref{eq:occupancy-entropy}
$\Box.$

\vspace{0.2cm}

\subsection{Monotone Submodularity \label{subsec:Proof-of-Submodularity}}

To prove Proposition\textbf{ \ref{prop:mutual_information_submodular}},
we first require the following result. \begin{lemma}\label{lemma:mi_submo}
$\forall\boldsymbol{R}\subseteq\boldsymbol{Z}\in\pmb{\mathbb{\mathcal{Z}}}$
and $\forall\boldsymbol{Y}\in\pmb{\mathcal{Z}}\setminus\boldsymbol{Z}$:
\begin{align*}
I(\boldsymbol{X};\boldsymbol{Z},\boldsymbol{Y})-I(\boldsymbol{X};\boldsymbol{Z}) & \leq I(\boldsymbol{X};\boldsymbol{R},\boldsymbol{Y})-I(\boldsymbol{X};\boldsymbol{R}).
\end{align*}
\textbf{Proof }: Since $\boldsymbol{R}\subseteq\boldsymbol{Z}\in\pmb{\mathbb{\mathcal{Z}}}$,
using the mutual information inequalities \cite[p.50]{cover2012elements},
we have: 
\begin{eqnarray}
 & \!\!\!\!\!\!\!\!\!\!\!\!\!\!\!\!\!\!\!\!\!\!\!\!\!\! & I(\boldsymbol{Z};\boldsymbol{Y})\geq I(\boldsymbol{R};\boldsymbol{Y})\nonumber \\
 & \Leftrightarrow & h(\boldsymbol{Y})-h(\boldsymbol{Y}|\boldsymbol{Z})\geq h(\boldsymbol{Y})-h(\boldsymbol{Y}|\boldsymbol{R})\\
 & \Leftrightarrow & h(\boldsymbol{Y}|\boldsymbol{R})\geq h(\boldsymbol{Y}|\boldsymbol{Z}),\nonumber \\
 & \Leftrightarrow & h(\boldsymbol{R},\boldsymbol{Y})-h(\boldsymbol{R})\geq h(\boldsymbol{Z},\boldsymbol{Y})-h(\boldsymbol{Z}).\label{eq_18}
\end{eqnarray}
Further, nothing that $I(\boldsymbol{Z};\boldsymbol{Y}|\boldsymbol{X})=I(\boldsymbol{R};\boldsymbol{Y}|\boldsymbol{X})=0$
because $\boldsymbol{Y}$ is independent of $\boldsymbol{R}$ and
$\boldsymbol{Z}$, given $\boldsymbol{X}$, we have, 
\begin{align*}
h(\boldsymbol{Y}|\boldsymbol{X}) & =h(\boldsymbol{Y}|\boldsymbol{X},\boldsymbol{Z})+I(\boldsymbol{Z};\boldsymbol{Y}|\boldsymbol{X})=h(\boldsymbol{Y}|\boldsymbol{X},\boldsymbol{Z})\\
 & =h(\boldsymbol{X},\boldsymbol{Z},\boldsymbol{Y})-h(\boldsymbol{X},\boldsymbol{Z}),\\
 & =h(\boldsymbol{Y}|\boldsymbol{X},\boldsymbol{R})+I(\boldsymbol{R};\boldsymbol{Y}|\boldsymbol{X})\\
 & =h(\boldsymbol{X},\boldsymbol{R},\boldsymbol{Y})-h(\boldsymbol{X},\boldsymbol{R}).
\end{align*}
and hence, 
\begin{align}
h(\boldsymbol{X},\boldsymbol{Z},\boldsymbol{Y})-h(\boldsymbol{X},\boldsymbol{Z})=h(\boldsymbol{X},\boldsymbol{R},\boldsymbol{Y})-h(\boldsymbol{X},\boldsymbol{R}).\label{eq_21}
\end{align}
Subtracting \eqref{eq_18} from \eqref{eq_21}, we have: 
\begin{align*}
 & \!\!\!\!\!\!\!\!\![h(\boldsymbol{X},\boldsymbol{Z},\boldsymbol{Y})-h(\boldsymbol{X},\boldsymbol{Z})]-[h(\boldsymbol{Z},\boldsymbol{Y})-h(\boldsymbol{Z})]\\
 & \geq[h(\boldsymbol{X},\boldsymbol{R},\boldsymbol{Y})-h(\boldsymbol{X},\boldsymbol{R})]-[h(\boldsymbol{R},\boldsymbol{Y})-h(\boldsymbol{R})]
\end{align*}
Note from the differential entropy chain rules \cite[p.253]{cover2012elements}
that $h(\boldsymbol{X}|\boldsymbol{Z},\boldsymbol{Y})=h(\boldsymbol{X},\boldsymbol{Z},\boldsymbol{Y})-h(\boldsymbol{Z},\boldsymbol{Y})$
and $h(\boldsymbol{X}|\boldsymbol{Z})=h(\boldsymbol{X},\boldsymbol{Z})-h(\boldsymbol{Z})$.
Substituting these into the above gives 
\begin{align*}
\!\!\!\!\!\!\!\!\! & h(\boldsymbol{X}|\boldsymbol{Z},\boldsymbol{Y})-h(\boldsymbol{X}|\boldsymbol{Z})\geq h(\boldsymbol{X}|\boldsymbol{R},\boldsymbol{Y})-h(\boldsymbol{X}|\boldsymbol{R})\\
\Leftrightarrow & [h(\boldsymbol{X})-h(\boldsymbol{X}|\boldsymbol{Z},\boldsymbol{Y})]-[h(\boldsymbol{X})-h(\boldsymbol{X}|\boldsymbol{Z})]\\
 & \leq[h(\boldsymbol{X})-h(\boldsymbol{X}|\boldsymbol{R},\boldsymbol{Y}]-[h(\boldsymbol{X})-h(\boldsymbol{X}|\boldsymbol{R})],\\
\Leftrightarrow & I(\boldsymbol{X};\boldsymbol{Z},\boldsymbol{Y})-I(\boldsymbol{X};\boldsymbol{Z})\leq I(\boldsymbol{X};\boldsymbol{R},\boldsymbol{Y})-I(\boldsymbol{X};\boldsymbol{R}).\mathbf{\Box.}
\end{align*}
\end{lemma}

\textbf{Proof of Proposition \ref{prop:mutual_information_submodular}:}
Denote $\boldsymbol{\breve{\pi}}(\boldsymbol{Z}_{1}\cup\boldsymbol{Z}_{2})\equiv\boldsymbol{\pi}(\boldsymbol{Z}_{1},\boldsymbol{Z}_{2})$,
then according to (3.53) and (25.36) in \cite{mahler2014advances}:
$\int\boldsymbol{\breve{\pi}}(\boldsymbol{Z}_{1}\cup\boldsymbol{Z}_{2})\delta(\boldsymbol{Z}_{1}\cup\boldsymbol{Z}_{2})=\int{\boldsymbol{\pi}(\boldsymbol{Z}_{1},\boldsymbol{Z}_{2})}\delta\boldsymbol{Z}_{1}\delta\boldsymbol{Z}_{2}$.
Further, noting that 
\begin{gather*}
I(\boldsymbol{X};\boldsymbol{Z})=\int\boldsymbol{\pi}(\boldsymbol{X,Z})\ln\left(\dfrac{\boldsymbol{\pi}(\boldsymbol{X,Z})}{\boldsymbol{\pi}(\boldsymbol{X})\boldsymbol{\pi}(\boldsymbol{Z})}\right)\delta\boldsymbol{X}\delta\boldsymbol{Z}\\
I(\boldsymbol{X};\boldsymbol{Z_{1},Z_{2}})=\int\boldsymbol{\pi}(\boldsymbol{X,Z_{1},Z_{2}})\ln\left(\dfrac{\boldsymbol{\pi}(\boldsymbol{X,Z_{1},Z_{2}})}{\boldsymbol{\pi}(\boldsymbol{X})\boldsymbol{\pi}(\boldsymbol{Z_{1},Z_{2}})}\right)\delta\boldsymbol{X}\delta\boldsymbol{Z_{1}}\delta\boldsymbol{Z_{2}}.
\end{gather*}
we have $I(\boldsymbol{X};\boldsymbol{Z}_{1}\cup\boldsymbol{Z}_{2})=I(\boldsymbol{X};\boldsymbol{Z}_{1},\boldsymbol{Z}_{2})$.

Due to the mutual exclusiveness of the multi-object measurements between
the agents, and the independence of an agent's measurement from the
other agent's actions, $\boldsymbol{Z}_{\boldsymbol{A},j}=\uplus_{\boldsymbol{a}\in\boldsymbol{A}}\boldsymbol{\boldsymbol{Z}}_{\boldsymbol{a},j}$,
where $\boldsymbol{\boldsymbol{Z}}_{(a^{(m)},m),j}$ is the measurement
set received by agent $m$ at time $j$ after it has taken action
$a^{(m)}$. Hence, for $\boldsymbol{A}\subseteq\boldsymbol{B}\subset\pmb{\mathbb{A}}$,
$\boldsymbol{a}\in\pmb{\mathbb{A}}\setminus\boldsymbol{B}$, we have:
$\boldsymbol{Z}_{\boldsymbol{A}}\subseteq\boldsymbol{Z}_{\boldsymbol{B}}$
and $\boldsymbol{Z}_{\boldsymbol{a}}\in\pmb{\mathcal{Z}}\setminus\boldsymbol{Z}_{\boldsymbol{B}}$.
Thus, it follows from Lemma \ref{lemma:mi_submo} that 
\begin{gather*}
I(\boldsymbol{X};\boldsymbol{Z}_{\boldsymbol{B}},\boldsymbol{Z}_{\boldsymbol{a}})-I(\boldsymbol{X};\boldsymbol{Z}_{\boldsymbol{B}})\leq I(\boldsymbol{X};\boldsymbol{Z}_{\boldsymbol{A}},\boldsymbol{Z}_{\boldsymbol{a}})-I(\boldsymbol{X};\boldsymbol{Z}_{\boldsymbol{A}}),
\end{gather*}
which is equivalent to 
\begin{gather*}
I(\boldsymbol{X};\boldsymbol{Z}_{\boldsymbol{B}\cup\{\boldsymbol{a}\}})-I(\boldsymbol{X};\boldsymbol{Z}_{\boldsymbol{B}})\leq I(\boldsymbol{X};\boldsymbol{Z}_{\boldsymbol{A}\cup\{\boldsymbol{a}\}}-I(\boldsymbol{X};\boldsymbol{Z}_{\boldsymbol{A}}).
\end{gather*}
i.e., $I(\boldsymbol{X};\boldsymbol{Z}_{\boldsymbol{A}})$ is a \textit{submodular}
set function. Further, using the chain rule we have: 
\begin{gather*}
I(\boldsymbol{X};\boldsymbol{Z}_{\boldsymbol{A}},\boldsymbol{Z}_{\boldsymbol{a}})-I(\boldsymbol{X};\boldsymbol{Z}_{\boldsymbol{A}})=I(\boldsymbol{X};\boldsymbol{Z}_{\boldsymbol{A}}|\boldsymbol{Z}_{\boldsymbol{a}})\geq0
\end{gather*}
i.e., $I(\boldsymbol{X};\boldsymbol{Z}_{\boldsymbol{A}})$ is a \textit{ monotone
submodular} set function~$\Box$.

\textbf{Proof of Proposition \ref{theorem:V2}: }Let $\boldsymbol{Y}_{\boldsymbol{A}}^{(i)}=\uplus_{\boldsymbol{a}\in\boldsymbol{A}}\boldsymbol{Y}_{\boldsymbol{a}}^{(i)}$,
where $\boldsymbol{Y}_{(a^{(m)},m)}^{(i)}=(Y_{(a^{(m)},m)}^{(i)},m)$,
and $Y_{(a^{(m)},m)}^{(i)}$ is the binary measurement $\delta_{\emptyset}[\boldsymbol{Z}_{(a^{(m)},m)}(\varkappa^{(i)})]$
received by agent $m$ from cell $\varkappa^{(i)}$, after it has
taken action $(a^{(m)},m)$. Note that the binary measurement $Y_{\boldsymbol{A}}^{(i)}=\delta_{\emptyset}[\boldsymbol{Z}_{\boldsymbol{A}}(\varkappa^{(i)})]$
received from cell $\varkappa^{(i)}$ can be written as $\prod_{\boldsymbol{a}\in\boldsymbol{A}}Y{}_{\boldsymbol{a}}^{(i)}$,
due to the independence of the agent's measurement from the other
agent's actions. We first show by induction that 
\begin{align}
h(O^{(i)}|\boldsymbol{Y}{}_{\boldsymbol{A}}^{(i)}) & =h(O^{(i)}|Y_{\boldsymbol{A}}^{(i)})\label{eq:56}
\end{align}
It is trivial to confirm that \eqref{eq:56} holds for $\{\boldsymbol{a}\}$.
Suppose that \eqref{eq:56} holds for a non-empty $\boldsymbol{A}$,
then 
\begin{gather}
h(O^{(i)}|\boldsymbol{Y}_{\boldsymbol{A}}^{(i)})=h(O^{(i)}|Y_{\boldsymbol{A}}^{(i)})=\textrm{Pr}(Y_{\boldsymbol{A}}^{(i)}=1)h(O^{(i)}|Y_{\boldsymbol{A}}^{(i)}=1).\label{eq:58}
\end{gather}
Decomposing $\boldsymbol{Y}_{\boldsymbol{A}\cup\{\boldsymbol{a}\}}^{(i)}$,
we have 
\begin{align}
 & h(O^{(i)}|\boldsymbol{Y}_{\boldsymbol{A}\cup\{\boldsymbol{a}\}}^{(i)})=h(O^{(i)}|\boldsymbol{Y}_{\boldsymbol{A}}^{(i)},\boldsymbol{Y}_{\boldsymbol{a}}^{(i)})\nonumber \\
 & =\textrm{Pr}(\boldsymbol{Y}_{\boldsymbol{a}}^{(i)}=0)h(O^{(i)}|\boldsymbol{Y}_{\boldsymbol{A}}^{(i)},\boldsymbol{Y}_{\boldsymbol{a}}^{(i)}=0)\nonumber \\
 & \quad\quad+\textrm{Pr}(\boldsymbol{Y}_{\boldsymbol{a}}^{(i)}=1)h(O^{(i)}|\boldsymbol{Y}_{\boldsymbol{A}}^{(i)},\boldsymbol{Y}_{\boldsymbol{a}}^{(i)}=1)\\
 & =\textrm{Pr}(\boldsymbol{Y}_{\boldsymbol{a}}^{(i)}=1)h(O^{(i)}|\boldsymbol{Y}_{\boldsymbol{A}}^{(i)},\boldsymbol{Y}_{\boldsymbol{a}}^{(i)}=1)\label{eq:26}
\end{align}
since $h(O^{(i)}|\boldsymbol{Y}_{\boldsymbol{A}}^{(i)},\boldsymbol{Y}_{\boldsymbol{a}}^{(i)}=0)=h(O^{(i)}|Y_{\boldsymbol{A}\cup\{\boldsymbol{a}\}}^{(i)}=0)=0$.
Substitute \eqref{eq:58} into \eqref{eq:26}, we have: 
\begin{align*}
 & h(O^{(i)}|\boldsymbol{Y}_{\boldsymbol{A}\cup\{\boldsymbol{a}\}}^{(i)})\\
 & =\textrm{Pr}(\boldsymbol{Y}_{\boldsymbol{a}}^{(i)}=1)\textrm{Pr}(Y_{\boldsymbol{A}}^{(i)}=1)h(O^{(i)}|Y_{\boldsymbol{A}}^{(i)}=1,\boldsymbol{Y}_{\boldsymbol{a}}^{(i)}=1)\\
 & =\textrm{Pr}(Y_{\boldsymbol{A}\cup\{\boldsymbol{a}\}}^{(i)}=1)h(O^{(i)}|Y_{\boldsymbol{A}\cup\{\boldsymbol{a}\}}^{(i)}=1)\\
 & =h(O^{(i)}|Y_{\boldsymbol{A}\cup\{\boldsymbol{a}\}}^{(i)}).\quad\quad\quad
\end{align*}

From Proposition~\ref{prop:mutual_information_submodular}, $-h(O^{(i)}|\boldsymbol{Y}_{\boldsymbol{A}}^{(i)})$,
and hence $-h(O^{(i)}|Y_{\boldsymbol{A}}^{(i)})$, are monotone submodular.
Consequently, it follows from \cite[pp.272]{nemhauser1978analysis}
that $V_{2}(\boldsymbol{A})$ is monotone submodular because it is
a positive linear combination $-h(O^{(i)}|Y_{\boldsymbol{A}}^{(i)})$
~$\Box$.
\end{document}